\documentclass{article}

\pdfoutput=1
\usepackage{arxiv}

\usepackage[utf8]{inputenc} 
\usepackage[T1]{fontenc}    
\usepackage{hyperref}       
\usepackage{url}            
\usepackage{booktabs}       
\usepackage{amsfonts}       
\usepackage{nicefrac}       
\usepackage{microtype}      
\usepackage{lipsum}
\usepackage{graphicx}
\usepackage[authoryear]{natbib}
\usepackage{newtxmath}
\usepackage{subeqnarray}
\graphicspath{ {./images/} }

\title{MHD Flow Regimes in Annular Channel}

\author{
 Kaiyu Zhang \\
  School of Astronautics\\
  Beihang University\\
  Beijing, 102206, China\\
  \texttt{zkyae@buaa.edu.cn} \\
   \And
 Yibai Wang \\
  School of Astronautics\\
  Beihang University\\
  Beijing, 102206, China\\
  \And
  Haibin Tang \\
  School of Space and Environment\\
   Beihang University\\
    Beijing, 102206, China\\
      \texttt{thb@buaa.edu.cn} \\
  \And
 Lijun Yang \\
  School of Astronautics\\
  Beihang University\\
  Beijing, 102206, China\\
}

\begin{document}
\maketitle
\begin{abstract}
One method and two results are contributed to the complete understanding about MHD laminar flow in annular channel with transverse magnetic field in this paper.
In terms of the method, a computationally cheap semi-analytic algorithm is developed based on spectral method and perturbation expansion.
By virtue of the fast computation, dense cases with almost continuous varying Hartmann number $M$, Reynolds number $Re$ and cross-section ratio $\eta$ are calculated to explore the flow patterns that are missed in previous research.
In terms of the results of inertialess regime, we establish the average velocity map and electric-flow coupling delimitation in $\eta$-$M$ space.
Seven phenomenological flow patterns and their analytical approaches are identified.
In terms of the results of inertial regime, we examine the law of decreasing order-of-magnitude of inertial perturbation on primary flow with increasing Hartmann number.
The proposed semi-analytic solution coincides with the $Re^2/M^{4}$ suppression theory of Baylis \& Hunt ({\it J. Fluid Mech.}, vol. 43, 1971, pp. 423-428) in the case of $M<40$.
When $M>40$, the pair of trapezoid vortices of secondary flow begins to crack and there is therefore a faster drop in inertial perturbation as $Re^2/M^{5}$, which is a new suppression theory.
When $M>80$, the anomalous reverse vortices are fully developed near Shercliff layers resulting in the weaker suppression mode of $Re^2/M^{2.5}$, which confirms the theoretical prediction of Tabeling \& Chabrerie ({\it J. Fluid Mech.}, vol. 103, 1981, pp. 225-239).
\end{abstract}


\section{Introduction}
\label{sec:introduction}
\begin{figure}
	\centerline{\includegraphics[width=0.5\textwidth]{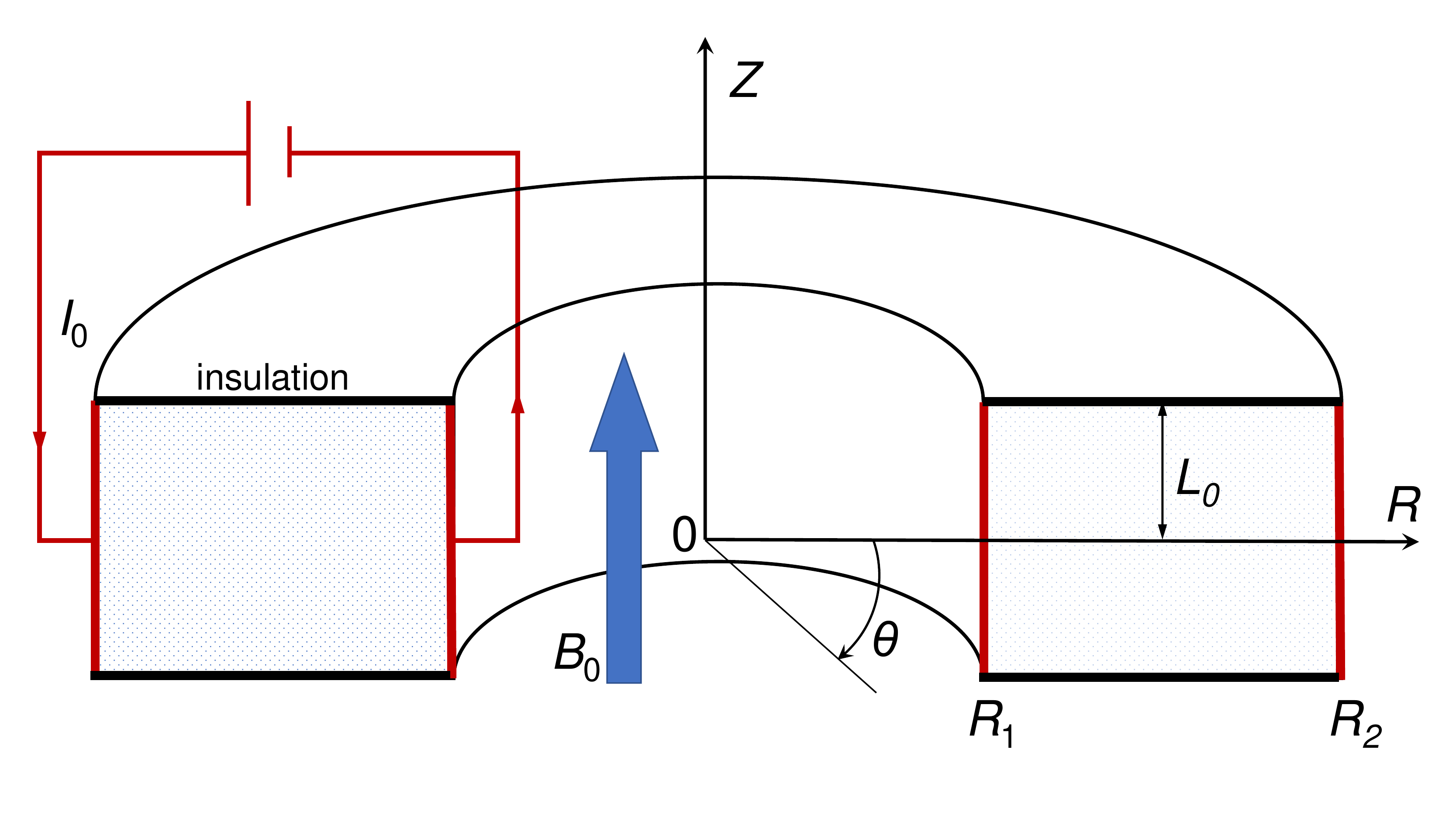}}
	\caption{Scheme of the MHD annular duct. The current investigation is on the $(r,z)$ plane.}
	\label{fig:apparatus}
\end{figure}

Magnetohydrodynamic (MHD) flow in the annular channel with transverse magnetic field is of great interest due to the experimental search for magneto-rotational-instability (MRI) in astrophysics \cite{balbus_instability_1998}, MHD dynamo in geophysics \cite{larmor1919possible}, and swirling actuators in engineering \cite{zhang_two-dimensional_2020}.
The idea of this apparatus, of which schematic is shown in figure \ref{fig:apparatus}, is using radial electric field and axial magnetic field to make azimuthal Lorenz forces and drive the azimuthal rotation of conductive fluids (primary flow).
At the same time, centrifugal force causes radial velocity transport, and then the secondary eddies emerge on the radial-axial plane (secondary flow).

The relevant research can be traced back to the experiment of \citet{baylis_detection_1964}, in which the onset of instability was detected through the drop in measured voltage.
However, at that time, it was unclear whether this onset point was caused by the enhancement of secondary flow or transition into turbulence.
Later, \citet{baylis_mhd_1971} established the first analytical solution to the velocity profile of the primary rotation flow at great Hartman number $M$ using boundary-layer-analysis method.
This research theoretically proved that increasing $M$ can suppress the magnitude of inertial force and secondary flow to an extent of $Re^2/M^{4}$, and thus stabilize the primary flow.
This $Re^2/M^{4}$ law has been commonly used to judge whether the secondary flow is coupled into the laminar-to-turbulence transition in later research.
\citet{tabeling_magnetohydrodynamic_1981} performed perturbation expansion of curvature of the annular channel and recalculated the secondary flow. 
His analytical study identified there will be the reverse vortexes adjacent to the conductive walls in secondary flow if the Hartmann number is great.
And the presence of reverse vortexes leads to a weaker suppression mode of inertial effect, which is $Re^2/M^{2.5}$.
Shortly afterwards, \citet{tabeling_sequence_1982} experimentally investigated the sequences of Taylor instability caused by centrifugal force and secondary flow before the onset of turbulence in the case of $M\sim10^2$. 
On the other hand, \citet{moresco_experimental_2004} conducted the experiment with by far higher Hartmann number ($M~10^3$) for the purpose of the secondary-flow-free turbulence transition.
This experiment indicated the critical parameter for the turbulent onset is $Re/M\approx380$.
Besides, the laminar part of experiment data showed the friction factor $F\approx 2M/Re$ for the case of square cross-section, which has been an important result used to verify the numerical or analytical laminar models.

\citet{khalzov_magnetorotational_2006} threw light on the global instability modes without dissipation effect by the method of spectral analysis, and found that the flow can be stable to the axisymmetric mode for certain geometric parameters of the device. 
Years later, \citet{khalzov_equilibrium_2010}, \citet{vantieghem_numerical_2011} and \citet{zhao_instabilities_2012} performed detailed numerical investigation.
Their numerical simulations observed the coexistence of Hartmann layers and Shercliff layers in primary flow and the morphology of toroidal vortices in secondary flow.
These three simulations came to a consistent conclusion that toroidal vortices of secondary flow can strengthen azimuthal velocity of primary flow on radially outer side, but weaken that on inner side.
Particularly, for the case with $M=100$ and $M=400$, \citet{vantieghem_numerical_2011} confirmed the presence of reverse vortexes adjacent to the conductive walls as predicted by \citet{tabeling_magnetohydrodynamic_1981}.
The Doppler velocimetry started to be used in the relevant research since the experiment of \citet{stelzer_experimental_2015-1}, in which the structure of electrodes was modified in the hope of observing Sheriff-layer-free flow.
With the Doppler velocimetry and the common electrodes, \citet{boisson_inertial_2017} experimentally confirmed the toroidal vortices that had been calculated by previous numerical simulations for the first time.
Furthermore, he identified that duct geometry has dramatic effect on the frequency of instability waves, which shed new light on the geometry effect, while all previous numerical research take the fixed cross-section of which high-width ratio $\eta$ is 1.
Recently, to extend our knowledge into large $\eta$ case, \citet{poye_scaling_2020} employed numerical simulation for much more cases with the varied cross-section shapes, and then identified fruitful scale laws for different flow structures. 
Specifically, he pinpointed that it is possible that Hartmann layers exist while Shercliff layers overlap when $\eta\gg1$, while the opposite case of $\eta\ll1$ was recommended as future study.

Against this background, it is true that a large number of experimental and numerical data have been available for this type of MHD channel flow.
But even the laminar case is far from thoroughly understood. 
There are at least three problems to be tackled before completing the research about the MHD annular laminar flow.
\begin{itemize}
	\item[1] \textbf{In inertialess regime, how many flow patterns there are?} the research of \citet{poye_scaling_2020} indicates that different geometric parameters and Hartmann numbers lead to the flow patterns, of which boundary layers are different from the analytical theory of \citet{baylis_mhd_1971}. 
	However, the operating conditions of previous simulations and experiments are limited.
	Most of the combination of geometric parameters and Hartmann numbers have never been reached.
	Moreover, it is still not certain that how many flow patterns there are and what features they have.
	\item[2] \textbf{In inertial regime, how the vortexes develop in secondary flow by degrees?}
	the simulation of \citet{khalzov_equilibrium_2010} show there is only one pair of vortexes in secondary flow with moderate Hartmann number $M=30$. 
	The simulation of \citet{zhao_instabilities_2012} show there are two pairs of vortexes when $M\approx100$. 
	The simulation of \citet{vantieghem_numerical_2011} show there will be thin reverse vortexes apart from the two pair of main vortexes when $M\approx400$. 
	However, because the values of Hartmann number in these simulations are scattered, it is still not clear how the vortexes develop in secondary flow as Hartmann number increases.
	And we do not know the exact $M$ criticality when vortexes break and reverse vortexes emerge neither.  
	\item[3] \textbf{In terms of the criteria of inertial and inertialess regime, which scale law shall be applied?}
	The estimate formula is important to the MRI research, because researchers prefer no effect of secondary flow on the profile of primary flow.
	The consensus among previous research is that increasing Hartmann number can suppress inertial effect. 
	The analytical analysis of \citet{baylis_mhd_1971} believe the order of $M$ in the suppression is $-4$, while \citet{tabeling_magnetohydrodynamic_1981} thought $M^{-4}$ theory underestimate the axial velocity near Shercliff layers and should be modified into $M^{-2.5}$. 
	However, most of later research applies $M^{-4}$ theory to the estimation of inertial effect, while $M^{-2.5}$ theory has not been given enough attention.
	The point is, to our best knowledge, there is neither numerical nor experimental data conforming either of the two theories.
	It seems that each of the two theories is supposed to have a specific sphere of application, which is still unknown to the scientific community.
\end{itemize}

It is clear that the solution to the three problems entails studying numerous physical cases with almost continuously changing physical and geometric parameters, which can hardly be achieved by existing research methods of numerical simulations.
In this paper, a novel semi-analytic method is proposed to meet the demands.  
The description of this algorithm is recorded in section \ref{sec:method}.
In section \ref{sec:inertialess}, we will try to use this new method to cover all existing experimental and numerical data for inertialess regime, and meanwhile conduct a comprehensive exploration of the flow patterns to answer the first question. 
Section \ref{sec:inertial} is devoted to the inertial regime.
We will harmonize the two disputed laws of inertial effect to answer the last two questions.

\section{Equations of Problem}
\label{sec:headings}
We consider the incompressible dissipative MHD fluid in the annular channel with a rectangular cross-section as shown in Fig.\ref{fig:apparatus}.
We set the cylindrical coordinate system $R,\theta, Z$ whose origin is at the symcenter of the toroid.
The outer radius and inner radius are $R_2$ and $R_1$, respectively.
The height of toroid is bisected by the $R$-$\theta$ plane into two halves with the length of $L_0$.
The symbol notations are presented in table \ref{tab:symbol}.
The walls at $R=R_1$ and $R=R_2$ are called side walls, which are conductive; the walls at $Z=\pm1$ are called end walls, which are insulative.
In this system, the external magnetic field $B_0$ is imposed axially.
And the current goes through the duct from outer to inner, inducing the azimuthal self magnetic fields.

This study assumes that:
\begin{itemize}
	\item the flow is laminar and steady.
	\item the steady field variables are axisymmetric, namely $\partial /\partial \theta \equiv0$ for any physical quantity.
	\item the induced magnetic field can be ignored compared with $B_0$ \citep{davidson_introduction_2016}. Mathematically, magnetic Reynolds number $Re_m\ll1$ and magnetic Prandtl number $Pr_m\ll1$.
\end{itemize}
\begin{table}
	\begin{center}	
		\def~{Mphantom{0}}
		\begin{tabular}{lll}
			Notions & Dimensional& Non-dimensional\\
			\hline	
			\multicolumn{3}{c}{constants}\\
			half of the duct height (scaling the length) & $L_0$ &   \\ 
			inner radius	& $R_1$ & $r_1\equiv R_1/L_0=(1-\kappa)/{\kappa\eta}$  \\
			outer radius& $R_2$ & $r_2\equiv R_2/L_0=(1+\kappa)/{\kappa\eta}  $\\
			mean radius& $\bar{R}\equiv(R_1+R_2)/2$ & $\bar{r}\equiv(r_1+r_2)/2=1/{\kappa\eta}  $  \\
			duct width& $\Delta R\equiv R_2-R_1$ & $\Delta r\equiv r_2-r_1=2/{\eta}$  \\
			permeability of vacuum&$\mu_{0}$  & \\
			viscosity coefficient &$\mu$   & \\
			density  &$\rho$ &\\
			electric conductivity  &$\sigma$ &\\
			total current through the duct &$I_0$  & \\
			strength of the impose axial magnetic field &$B_0$  & \\
			scale of the velocity &$V_{0}\equiv{I_0}/(4\pi L_0\sqrt{\sigma\mu})$   & \\
			scale of the current density &$J_{0} \equiv{I_0}/(4\pi L_0^2)$  & \\
			\hline	
			\multicolumn{3}{c}{field variables}\\
			
			radial coordinate & $R$ & $r\equiv R/L_0$  \\
			axial coordinate & $Z$ & $z\equiv Z/L_0$\\
			vector of current density & $\boldsymbol{J}$&  $\boldsymbol{j}\equiv{\boldsymbol{J}}/J_{0}$  \\
			vector of induced magnetic field & $\boldsymbol{B}$ & $\boldsymbol{b}\equiv{\boldsymbol{B}}/(\mu_0J_{0}L_0)$\\
			vector of velocity &  $\boldsymbol{V}$ & $\boldsymbol{v} \equiv {\boldsymbol{V}}/V_0$   \\
			poloidal stream functions of velocity & & $w$\\
			angular momentum of the fluid & & $u\equiv rv_{\theta}$\\
			poloidal stream functions of current density & & $h\equiv rb_{\theta}$\\
			\hline
			\multicolumn{3}{c}{characterized numbers}\\
			integrate-average azimuthal velocity &$\bar{V}=V_0 \bar{v}$& $\bar{v} \equiv \int_{-1}^{1} \int_{r_1}^{r_2} {v_\theta} \text{d} r \text{d} z/2(r_1-r_2)$  \\
			ratio of inner to outer radius&  & $\kappa=(R_2-R_1)/(R_1+R_2)$  \\
			ratio of height to width&  & $\eta=2L_0/(R_2-R_1)$  \\
			Hartmann number &  & $M\equiv B_0 L_0\sqrt{\sigma/\mu}$  \\
			Reynolds number &  & $Re \equiv \rho V_0 L_0/\mu$  \\
			magnetic Reynolds number &  & $Re_m \equiv \mu_0 \sigma V_0 L_0$  \\
			magnetic Prandtl number &  & $Pr_m \equiv \mu_0 \sigma \mu/\rho$  \\
			\hline
		\end{tabular}
		\caption{Symbol notations.}
		\label{tab:symbol}
	\end{center}
\end{table}
With the above assumptions, the dimensional incompressible dissipative MHD equations are written as follows:
\begin{subeqnarray}
	\boldsymbol{\nabla\cdot V}&=&0,\\
	\boldsymbol{\nabla\cdot J}&=&0,\\
	\rho(\boldsymbol{V\cdot\nabla})\boldsymbol{V}&=&-\nabla p + \boldsymbol{J}\times\boldsymbol{B}_0+\mu{\nabla}^2\boldsymbol{V},\\
	\boldsymbol{\nabla}\times\boldsymbol{J}/\sigma&=&\boldsymbol{\nabla}\times(\boldsymbol{V}\times\boldsymbol{B}_0).\label{eqn:ohm law}
\end{subeqnarray}
Nondimensionalize length, velocity, current density by the scale of  $L_0$,  $V_{0}\equiv{I_0}/(4\pi L_0\sqrt{\sigma\mu})$, 
$J_{0} \equiv{I_0}/(4\pi L_0^2)$, respectively, as follows:
\begin{equation}
	r=R/L_0,\quad z=Z/L_0,\quad \boldsymbol{v}=\boldsymbol{V}/V_0, \quad \boldsymbol{j}=\boldsymbol{J}/J_0.
\end{equation}
In terms of $\boldsymbol{v}$, we define $u$ as the angular momentum and $w$ as stream functions of the velocity. Thanks to the axisymmetric assumption, we get:
\begin{equation}
	v_r=-\frac{\partial w}{r\partial z}, \quad v_\theta=\frac{u}{r}, \quad v_z=\frac{\partial w}{r\partial r}.
\end{equation}
In terms of $\boldsymbol{j}$, we define $h$ as stream functions of the current density. Based on the induction-less assumption combined with equation\ref{eqn:ohm law} b, we get:
\begin{equation}
	j_r=-\frac{\partial h}{r\partial z}, \quad j_\theta=M\frac{\partial w}{r \partial z}, \quad j_z=\frac{\partial h}{r\partial r}.
\end{equation}
In this context, the original MHD equations are transformed into:
\begin{subeqnarray}
	0&=&\Delta^*u+M\frac{\partial h}{\partial z}+Re\frac{u\otimes w}{r},\label{eqn:iu}\\
	0&=&\Delta^*h+M\frac{\partial u}{\partial z},\label{eqn:ih}\\
	0&=&\Delta^*\Delta^*w-M^2\frac{\partial^2 w}{\partial z^2}-r Re\left(w\otimes \frac{\Delta^*w}{r^2}+\frac{1}{r^3}\frac{\partial u^2}{\partial z}\right),\label{eqn:w}
\end{subeqnarray}
where the differential operators are defined as:
\refstepcounter{equation}
$$
f_1\otimes f_2 \equiv \frac{\partial f_1}{\partial r}\frac{\partial f_2}{\partial z}-\frac{\partial f_1}{\partial z}\frac{\partial f_2}{\partial r},\quad
\Delta^*=\frac{\partial^2}{\partial r^2}-\frac{1}{r}\frac{\partial}{\partial r}+\frac{\partial^2}{\partial z^2}
\eqno{(\theequation{\mathit{a},\mathit{b}})}
$$
Hartmann number $M$ and Reynolds number $Re$:
\refstepcounter{equation}
$$
M\equiv B_0 L_0\sqrt{\frac{\sigma}{\mu}}, \quad
Re\equiv \frac{\rho V_0L_0}{\mu}
\eqno{(\theequation{\mathit{a},\mathit{b}})}
\label{eqn:number}
$$
There are many parameters controlling the solution as listed in \ref{tab:symbol}. 
But in nondimensional case, the parameters can be deduced into only two physical number, namely Hartmann number $M$ and Reynolds number $Re$, and two geometric parameters of
\refstepcounter{equation}
$$
\kappa \equiv\frac{R_2-R_1}{R_2+R_1}, \quad
\eta \equiv\frac{2L_0}{R_2-R_1}
\eqno{(\theequation{\mathit{a},\mathit{b}})}.
\label{eqn:geometric}
$$
Specifically, $\kappa$ measures the curvature of annular channel; when $\kappa\rightarrow 0$, the channel can be treated as a straight one. 
$\eta$ refers to the ratio of height to width; the cross-section of channel is flat when $\eta\ll 1$, and tall when $\eta\gg 1$.
The nondimensional length can be expressed by $\kappa$ and $\eta$: $r_1=R_1/L_0=(1-\kappa)/{\kappa\eta}$, $r_2=R_2/L_0=(1+\kappa)/{\kappa\eta}$.
And the boundary conditions can be written as:
\begin{equation}\label{eqn:BC}
	\left\{
	\begin{array}{lll}
		r=(1\pm\kappa)/{\kappa\eta}: & u=w=\partial_r w=0,& \partial_r h=0\\
		z=\pm 1:& u=w=\partial_z w=0,& h=\pm 1
	\end{array}\right.
\end{equation}

\section{Semi-Analytical Algorithm}
\label{sec:method}
\subsection{Perturbation expansion of inertial force}
Expand the field variables as the series of $Re$:
\begin{subeqnarray}\label{eqn:pertubation}
	u&=&u_0 + u_1Re +u_2Re^2 +...u_nRe^n +... \\
	h&=&(z+h_0) + h_1Re +h_2Re^2 +...h_nRe^n +...\\
	w&=&w_0 + w_1Re +w_2Re^2 +...w_nRe^n +...
\end{subeqnarray}
We write $z+h_0$ rather than $h_0$ in equation \ref{eqn:pertubation}b so that the boundary conditions of $u_n, h_n, w_n$ are all homogeneous.
Submitting the expansion form to the original equation \ref{eqn:iu} produce the equations regards to each power of $Re$:
\begin{equation}\label{eqn:pm0}
	Re^0: \left\{
	\begin{array}{l}
		\begin{bmatrix}
			\Delta^* & M\partial_z\\
			M\partial_z & \Delta^* 
		\end{bmatrix}
		\begin{bmatrix}
			u_0\\
			h_0
		\end{bmatrix} = 
		\begin{bmatrix}
			-M\\
			0
		\end{bmatrix}	\\
		w_0=0
	\end{array}\right.
\end{equation}
\begin{equation}\label{eqn:pm1}
	Re^1: \left\{
	\begin{array}{l}
		u_1=h_1=0 \\
		(\Delta^*\Delta^*-M^2\frac{\partial^2 }{\partial z^2})w_1=\frac{1}{r^2}\frac{\partial u_0^2}{\partial z}
	\end{array}\right.
\end{equation}
\begin{equation}\label{eqn:pm2}
	Re^2: \left\{
	\begin{array}{l}
		\begin{bmatrix}
			\Delta^* & M\partial_z\\
			M\partial_z & \Delta^* 
		\end{bmatrix} 
		\begin{bmatrix}
			u_2\\
			h_2
		\end{bmatrix} = 
		-\frac{1}{r}
		\begin{bmatrix}
			u_0\otimes w_1\\
			0
		\end{bmatrix}	\\
		w_2=0
	\end{array}\right.
\end{equation}
We can solve $u_0 \& h_0$, $w_1$, $u_2 \& h_2$ et al. in turn and write down more equations regards to higher power of $Re$.
This process uncouples the primary flow $u$ and secondary flow $w$, though the primary flow $u$ is coupled with electromagnetic field $h$. 
In each term of $Re$, the original problem is transformed into the 2-order coupling PDEs for primary flow and electromagnetic field ($u \& h$ field):
\begin{subeqnarray}
		\Delta^*u+M\partial_z h&=&S_w\\ \label{eqn:uh-u}
		\Delta^*h+M\partial_z u&=&0\label{eqn:uh-h}
\end{subeqnarray}
and the 4-order PDE for secondary flow field ($w$ field):
\begin{equation} \label{eqn:pw}
	(\Delta^*\Delta^*-M^2\partial_{zz})w=S_u,
\end{equation}
where the sources $S_w$ and $S_u$ on the right side are obtained via the term with lower-order $Re$ as shown in equations \ref{eqn:pm0}, \ref{eqn:pm1} and  \ref{eqn:pm2}, and thus have been uncoupled with the left hand.
With this perturbation expansion, we proceed to solve equations \ref{eqn:uh-u} and equation \ref{eqn:pw} in subsection \ref{sec:s-m} and \ref{sec:p-s}, respectively.
\subsection{Spectral-Galerkin scheme for primary flow and electromagnetic field}\label{sec:s-m}
We try to find the weak solution of equations \ref{eqn:uh-u} by the Galerkin approach based on spectral method, and set:
\begin{subeqnarray}\label{eqn:semi-series}
	u&=\sum_{n=1}^{\infty}u_{r,n}(r)u_{z,n}(z),\\
	h&=\sum_{n=1}^{\infty}h_{r,n}(r)h_{z,n}(z)
\end{subeqnarray}
$u_{r,n}(r)$ and $h_{r,n}(r)$ are spectral-Bessel functions determined by the left side of equation\ref{eqn:uh-u} as:
\begin{equation}\label{eqn:spectral-Bessel}
	\frac{\partial^2 u_{r,n}}{\partial r^2}-\frac{1}{r}\frac{\partial u_{r,n}}{\partial r}=-\eta^2\alpha_n^2u_{r,n}, \quad  \frac{\partial^2h_{r,n}}{\partial r^2}-\frac{1}{r}\frac{\partial h_{r,n}}{\partial r}=-\eta^2\beta_n^2h_{r,n}
\end{equation}
where $\alpha_n$ and $\beta_n$ are the eigenvalues.
Appendix A shows the details for how $u_{r,n}(r)$ and $h_{r,n}(r)$ are deduced.
Define the integral operator for arbitrary functions $f_1$ and $f_2$ as:
\begin{equation}
	\langle f_1,f_2\rangle\equiv\int_{r_1}^{r_2}f_1f_2\frac{1}{r}\text{d} r
\end{equation}
The orthogonal law of Sturm-Liouville theory\citep{birkhoff1973source} tells  
\begin{equation}
	\langle  u_{r,i}, u_{r,j}\rangle=\langle  h_{r,i}, h_{r,j}\rangle=\delta_{i,j},
\end{equation}
in which $\delta$ is the Dirac function. With this orthogonality, integrating the product of $u_{r,i}$ and equation\ref{eqn:uh-u}a (the weight function is $1/r$) \citep{duncan1848principles} we can obtain the weak-form of equation\ref{eqn:uh-u}a
\begin{equation}
	(u''_{z,i}-\eta^2\alpha_i^2u_{z,i})+M \langle u_{r,i},h_{r,j}\rangle h'_{z,j}=\langle S_w,u_{r,i}\rangle\label{eqn:axial u}.
\end{equation}
$i$ is a specific positive integer and $j$ is supposed to follow Einstein summation convention.
Similarly, weak-form of equation\ref{eqn:uh-u}b is.
\begin{equation}
 	\langle u_{r,i},h_{r,j}\rangle(h''_{z,j}-\eta^2\beta_i^2h_{z,j})+M u'_{z,i}=0\label{eqn:axial h}
\end{equation}
At this point, the coupled two PDEs \ref{eqn:uh-u} has been reduced into a set of ODEs.
The detailed solution to them is presented in Appendix B.
It is noticed that the calculating cost has also been reduced dramatically compared with the common numerical scheme of finite difference,
To be specific, if the $n\times n$ resolution ($n^2$ meshing points) is required, only the magnitude of $n$ linear equations have to be solved by this spectral scheme, which takes on a global approach\citep{gottlieb1977numerical}.

\subsection{Pseudo-spectral scheme for secondary flow} \label{sec:p-s}
We take the tentative weak solution to equation \ref{eqn:w} as
\begin{equation}\label{eqn:pseudospectral}
	w(r,z)=\sum_{i}\sum_{j}c_{i,j}w_{r,i}(r)w_{z,j}(z)
\end{equation}
where $w_{r,i}$ and $w_{z,j}$ are the spectral functions, and $c_{i,j}$ is undetermined coefficients.
To meet the boundary condition \ref{eqn:BC}, the radial spectral functions for $w$ are of Bessel type as:
\begin{equation}
	w_{r,i}(r)=(r-r_1)(r_2-r)u_{r,i}(r)
\end{equation}
and the axial spectral functions are chosen as:
\begin{equation}
	w_{r,j}(z)=(1-z^2)\sin(j \pi z)
\end{equation}
Submitting the tentative solution \ref{eqn:pseudospectral} into the original equation \ref{eqn:w} at the given grid, we obtain the linear equation set of $c_{i,j}$.

\section{Results: Inertialess Regime}
\label{sec:inertialess}
In inertialess regime, the secondary flow can be neglected and then the velocity is along the poloidal coordinate.
This situation can exist in both high-$M$ or low-$M$ cases as long as $Re$ is small enough as equation \ref{eqn:iu} implies.
The equations of $u \& h$ field is:
\begin{subeqnarray}\label{eqn:uh-il}
	\Delta^*u+M\partial_z h=0\\
	\Delta^*h+M\partial_z u=0
\end{subeqnarray}
accompanied by the boundary condition \ref{eqn:BC}.
And the final solution $v(r,z)$ rests with $\kappa$, $\eta$, and $M$ exclusively.
By virtue of the computationally cheap semi-analytical algorithm proposed in section \ref{sec:method}, a large number ($200 \times 200=40,000$) of cases are carried out in the whole parameter space spanned by $\eta$ and $M$, each of which ranges from $10^{-2}$ to $10^{2}$.

Before investigating the microcosmic flow structures, we view the macroscopic flow characters subjected to different Hartmann numbers and geometric parameters.
The nondimensional integrate-average azimuthal velocity $\bar{v}$ is used to verify solutions and classify flow patterns \citep{poye_scaling_2020}:
\begin{equation}\label{eqn:average-v}
	\bar{v} = \frac{1}{2(r_1-r_2)}\int_{-1}^{1} \int_{r_1}^{r_2} {v_\theta} \text{d} r \text{d} z,
\end{equation}
According to equation \ref{eqn:average-v}, $\bar{v}$ is obtained by integrating the 2-dimensional azimuthal velocity $v$ from the calculated cases with different $\eta$ and $M$; the results are presented in figure \ref{fig:map} (a). 
It is noticed that there are phenomenological areas with different variational tends in this $\bar{v}$ contour figure.
Interestingly, these areas can be well delimited by the red, blue, and green curves with special physical meanings.
We name these areas a combination of capital letters C (coupled), D (decoupled), R (R-direction boundary layers), Z (Z-direction boundary layers) and N (no boundary layers).
The principle of delimiting is as followings. 
\subsection{Division of flow patterns}
\subsubsection{Electrically coupled modes}
These areas in figure \ref{fig:map} (a) fall into two classes C-areas (coupled) and D-areas (decoupled) by the red curve. 
It can be seen in figure \ref{fig:map} (a) that D areas see symmetrical trends of $\bar{v}$, of which contours are controlled by $\lg{M}\pm\lg{\eta}$.
By contrast, symmetry vanishes in C-areas where variation of $\bar{v}$ with $M$ levels off.
Further study indicates that this discrepancy is related to electrically coupling strength in the MHD system \ref{eqn:uh-il}, namely the extent to which the flow field $u$ disturbs the electromagnetic field $h$.
The totally decoupled state that $h$ is not influenced by $u$, corresponds to the case that $M\partial_z u$ is negligible in equation \ref{eqn:uh-il}b:
\begin{equation} \label{eqn:h-nc}
	\frac{\partial^2 h}{\partial r^2}-\frac{1}{r}\frac{\partial h}{\partial r}+\frac{\partial^2h}{\partial z^2}=0
\end{equation}
Given boundary condition \ref{eqn:BC}, equation \ref{eqn:h-nc} leads to a simple solution $h(r,z)=z$. 
To estimate the deviation from this decoupled state, we define the electrically coupling factor $\epsilon$ as
\begin{equation}\label{eqn:definite-ep} 
	\epsilon=\frac{||h(r,z)-z||}{||z||},
\end{equation}
where the energy norm operator is 
\begin{equation} \label{eqn:energy norm}
	||f(r,z)||\equiv\sqrt{\int_{-1}^{1}\int_{r_1}^{r_2}f(r,z)^2\frac{1}{r}\text{d} r \text{d} z}.
\end{equation}

$\epsilon$ is calculated in $\eta$-$M$ space; results are shown in figure \ref{fig:map} (b).
It can be seen that $\epsilon$ has different scale laws at small $\eta$ and large $\eta$.
For an analytical investigative, taking the first two terms in semi-analytical operation \ref{eqn:semi-series} we can get a analytical approximate solution for $h$:
\begin{equation}
h=z-M^2\frac{z-\text{csch}\left(\sqrt{\alpha _1 \eta ^2+M^2}\right) \sinh \left(z \sqrt{\alpha _1 \eta ^2+M^2}\right)}{\alpha _1 \eta ^2 \sqrt{\alpha _1 \eta ^2+M^2} \coth \left(\sqrt{\alpha _1 \eta ^2+M^2}\right)+M^2}.
\end{equation}
With the above analytical formula, $\epsilon$ can be readily obtained:
\begin{equation} \label{eqn:epm}
	\epsilon=\frac{M^2 \sqrt{\frac{6}{\alpha _1 \eta ^2+M^2}-\frac{9 \coth \sqrt{\alpha _1 \eta ^2+M^2}}{2 \sqrt{\alpha _1 \eta ^2+M^2}}-\frac{3}{2} \text{csch}^2\sqrt{\alpha _1 \eta ^2+M^2}+1}}{\alpha _1 \eta ^2 \sqrt{\alpha _1 \eta ^2+M^2} \coth \sqrt{\alpha _1 \eta ^2+M^2}+M^2}
\end{equation}
Complex as this expression is, it is controlled by three parameters $M$, $\eta$ and $\alpha_1$.
$\kappa$ is involved in $\epsilon$ through $\alpha_1$. 
However, mathematically, $\alpha_1$ is the first eigenvalue ranging from $\pi/2$ to $2$, and thus has marginal effect on $\epsilon$.
At this point, $\kappa$, namely the curvature of annular channel, can NOT determine the magnitude of electrically coupled factor $\epsilon$.
By contrast, the effect of $M$ and $\eta$ is significant, as can be seen in both equation \ref{eqn:epm} and figure\ref{fig:map} b.
Interestingly, the scale law of $\epsilon$ vs. $M$ is not obvious, but varies with $\eta$.
When $\eta\ll1$ and $M\ll1$, we have the Taylor's expansion for $\epsilon$ as
\begin{equation}\label{eqn:ep1}
	\epsilon=\frac{1}{3} \sqrt{\frac{2}{35}} M^2-\frac{13 \alpha _1 M^2}{45 \sqrt{70}}O(\eta ^2)\approx\frac{1}{3} \sqrt{\frac{2}{35}} M^2,
\end{equation}
therefore the contours of $\epsilon$ is approximately horizontal when $\eta\ll 1$ in $M$-$\eta$ space, as can be seen in figure \ref{fig:map} (b).
When $\eta\gg1$, the series is expanded as:
\begin{equation}\label{eqn:ep2}
	\epsilon=\frac{M/\eta^2}{\alpha _1+M/\eta^2}-\frac{\alpha _1^2}{2 \left(\alpha _1+M/\eta^2\right)^2 M/\eta^2}O(\frac{1}{\eta ^2})\approx\frac{M/\eta^2}{\alpha _1+M/\eta^2},
\end{equation}
therefore the slope of contours of $\epsilon$ is approximately 2 when $\eta\gg 1$ in $M$-$\eta$ space, as shown in figure \ref{fig:map} (b).

\subsubsection{Boundary layer modes}
\begin{figure}
	\centerline{\includegraphics[width=1\textwidth]{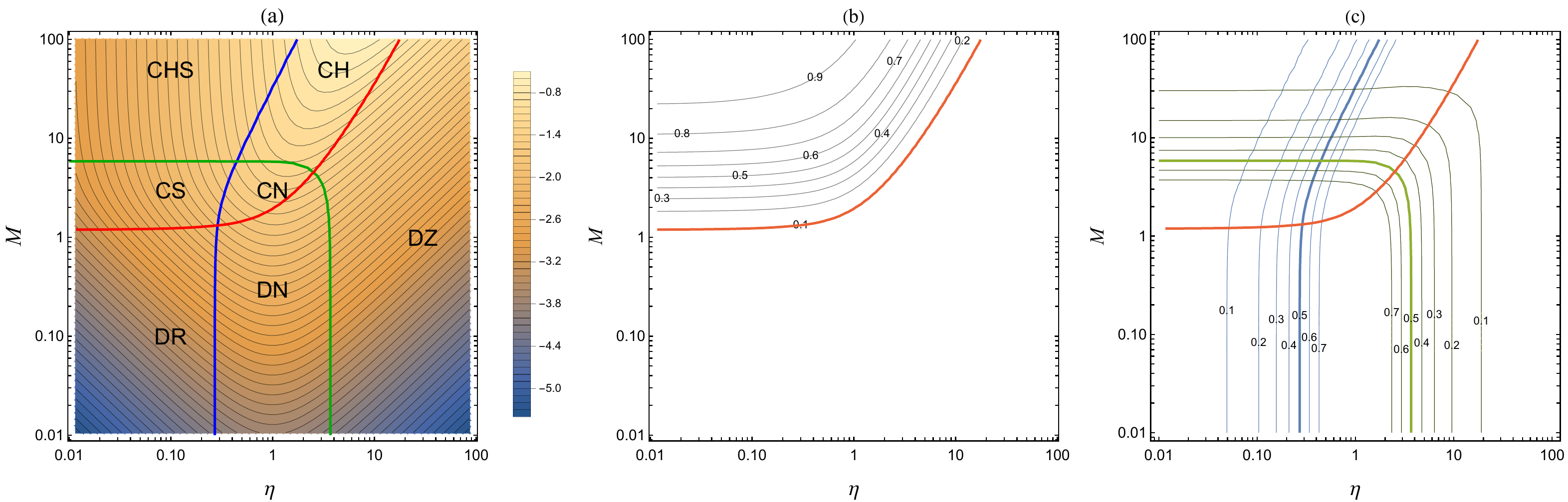}}
	\caption{Contours in the $\eta$-$M$ space with $\kappa=1/9$: (a) average velocity $\bar{v}$; (b) electrically coupling factor $\epsilon$; (c) thickness of boundary layer adjacent to side walls $\delta_r$ (blue) and thickness of boundary layer adjacent to end walls $\delta_z$ (green).
	The red thick curve in (a)\&(b) ($\epsilon=0.1$)  is the critical contour separating electrically coupled region (C-areas) and decoupled region (D-areas). The blue thick curve ($\delta_r=0.5$) and green thick curve ($\delta_z=0.5$) in (a)\&(c) are demarcation of regions with clear boundary layers.}
	\label{fig:map}
\end{figure}

The flow patters can also be identified from the perspective of boundary layers.
In this paper, the thickness of boundary layers are defined as the position where velocity reaches 95\% of its maximum
(this percentage can also be chosen as 90\% or 99\% without significantly different consequences).
Figure \ref{fig:map} (c) and (d) depict the R-direction (radial) thickness of layers $\delta_R$ and Z-direction (axial) thickness $\delta_Z$, respectively.
Particularly, layers with $\delta_R$ are adjacent to side walls while layers with $\delta_Z$ are adjacent to end walls.
They are modified by half of the radial dimension ($\Delta R/2$) and half of the axial dimension ($L$), respectively. 
Hence, $\delta_{R}$ or $\delta_Z \sim1$ means the opposite layers overlap, in other words, there is no clear boundary layers along this direction.
It is remarkable in figure \ref{fig:map} (c) and (d) that the thicknesses of boundary layers present distinct laws of variation in C-areas (above the red curve) and D-areas (below the red curve).

For the coupled case, the law of boundary layer is clearly understood by literature.
Early, \citet{shercliff_steady_1953} employed the singular perturbation method to the MHD flow in straight channel with transverse magnetic fields. 
They identified there were Shercliff layers adjacent to conductive walls with thickness of $~1/\sqrt{M}$\, and Hartmann layers adjacent to insulative walls with thickness of $~1/{M}$ under strong magnetic fields.
The layer thickness criteria were later validated and adopted in annular cases \citep{baylis_mhd_1971,tabeling_magnetohydrodynamic_1981,khalzov_equilibrium_2010}.
Considering that the nondimensional scales of radial and axial length of channel are $\sim1/\eta$ and $\sim1$, respectively, the modified thickness shall be
\begin{subeqnarray}\label{eqn:h layer}
	\delta_R\sim \eta/\sqrt{M},\\
	\delta_Z\sim1/{M}.
\end{subeqnarray}
These two formulas can well explain the diagonal blue line and horizontal green line in figure \ref{fig:map} (c), respectively.

For the decoupled case, little previous research is available.
Since $h=z$ is accepted in the decoupled case, the equation of $u$ can be reduced to 
\begin{equation}\label{eqn:uuc}
	\frac{\partial^2 u}{\partial r^2}-\frac{1}{r}\frac{\partial u}{\partial r}+\frac{\partial^2 u}{\partial z^2}+M=0
\end{equation} 
A closer view to the equation shows that Hartmann number $M$ do not influence the flow field structure, but merely scales up the magnitude of velocity.
That account for the vertical contours of thickness in $M$-$\eta$ space in figure \ref{fig:map} (c). 
To clarify the effect of $\eta$, we set $x=r \eta$ so that the boundary conditions $x=(1\pm\kappa)/\kappa$ are free of $\eta$, and then get:
\begin{equation}\label{eqn:uucx}
	\eta^2\left(\frac{\partial^2 u}{\partial x^2}-\frac{1}{x}\frac{\partial u}{\partial x}\right)+\frac{\partial^2 u}{\partial z^2}+M=0
\end{equation} 
The scale laws of thickness of boundary layer can be easily acquired from the above equation: when $\delta_R$ exists, the scales within radial layers are
\begin{equation}
	\eta^2\left(\frac{\partial^2 u}{\partial x^2}-\frac{1}{x}\frac{\partial u}{\partial x}\right)\sim\frac{\eta^2u}{\delta_R^2}, \quad \frac{\partial^2 u}{\partial z^2}\sim u, \quad\text{thus}\quad \delta_R\sim \eta;
\end{equation} 
when $\delta_Z$ exists, the scales within axial layers are
\begin{equation}
	\eta^2\left(\frac{\partial^2 u}{\partial x^2}-\frac{1}{x}\frac{\partial u}{\partial x}\right)\sim\eta^2u, \quad \frac{\partial^2 u}{\partial z^2}\sim \frac{u}{\delta_Z^2}, \quad\text{thus}\quad \delta_Z\sim \frac{1}{\eta}.
\end{equation} 
The scale laws $\delta_R\sim \eta$ and $\delta_Z\sim {1}/{\eta}$ account for the symmetry of contours of $\delta_R$ and $\delta_Z$ in $M$-$\eta$ space as can be seen in figure \ref{fig:map} (c).

These `boundary layers' in decoupled mode is fundamentally different from those in coupled mode.
The latter is caused by $M$ in magnetohydrodynamic context, while the former can be considered as a hydrodynamic profile stretched by coordinates.

\begin{figure}
	\centerline{\includegraphics[width=0.5\textwidth]{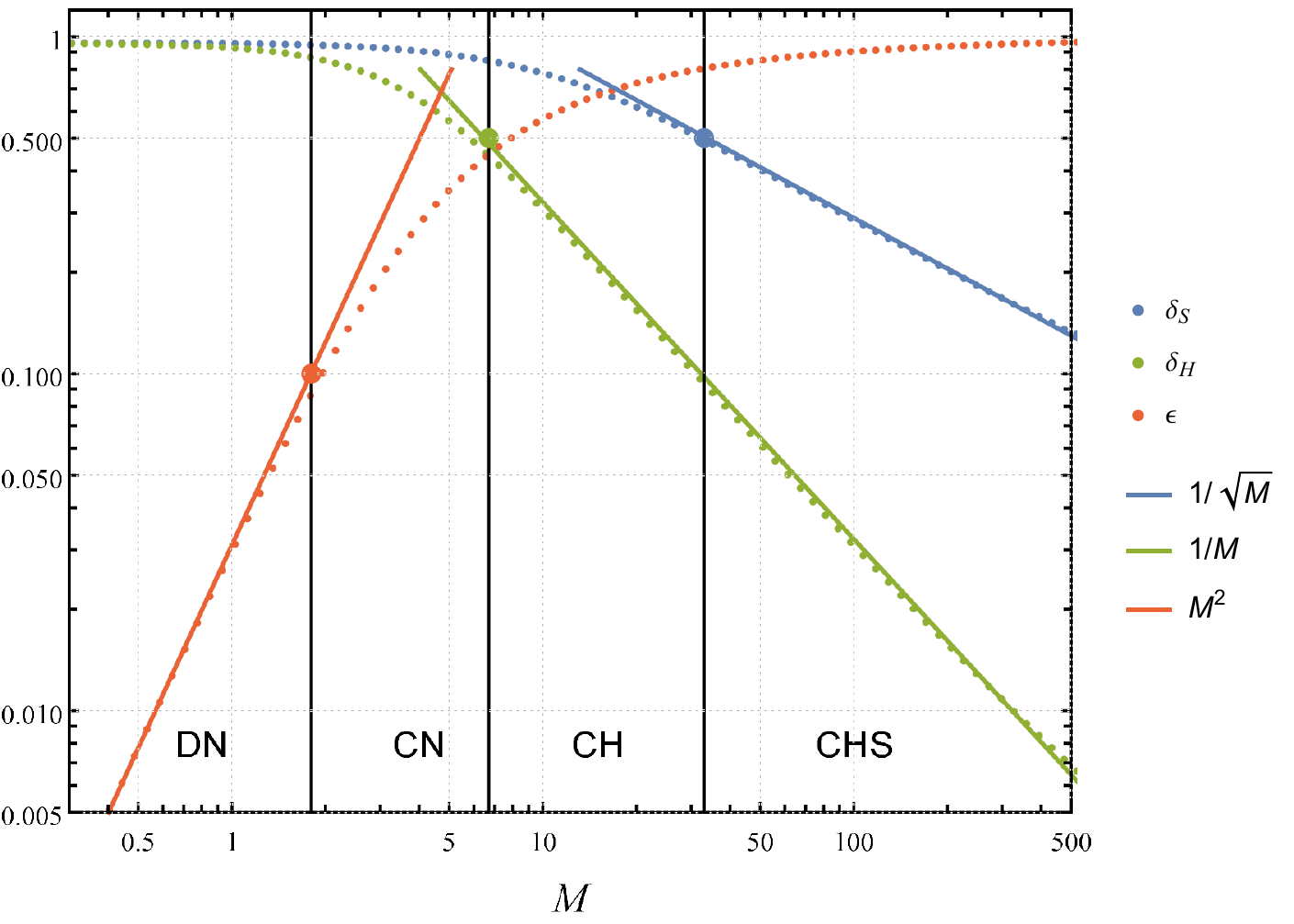}}
	\caption{Variations of the electrically coupling factor $\epsilon$ and the thickness of boundary layers $\delta_r$ and $\delta_r$ with increasing $M$ at $\eta=1, \kappa=1/9$. The dots are semi-analytical solution, while red, blue and green lines come from analytical theories \ref{eqn:ep1} and \ref{eqn:h layer} (a) and \ref{eqn:h layer} (b) respectively. The large points and plumb black lines mark the demarcations between different patterns.}
	\label{fig:de}
\end{figure}

\subsubsection{Analysis of criticality}
The red, blue and green curves, which divide areas in figure \ref{fig:map} (a), are $\epsilon=0.1$ contour in (b) and $\delta_R=0.5$, $\delta_Z=0.5$ contours in (c), respectively.
Although $0.1$ and $0.5$ are not exactly-obtained numerical indicators, these separating curves are not arbitrary at all.
Firstly, in terms of the red curve, as can be seen in figure \ref{fig:map} (c), the contours of $\delta_R$ and $\delta_Z$ see clear transition in the vicinity of $\epsilon=0.1$ contour.
Secondly, in terms of the blue and green curves, the contours of $\delta_R=0.5$ and $\delta_Z=0.5$ can well divide the contours of $\epsilon$ into horizontal, crooked and sloping section.
Finally, we depict the variation of $\epsilon$, $\delta_Z$, $\delta_R$ with $M$ at $\eta=1$ in figure \ref{fig:de}.
This figure illustrates that with the increase of $M$, $\epsilon$ (red dots) deviate from $\sim M^2$ law (red line) indicated by equation \ref{eqn:ep1}  in the vicinity of $\epsilon=1$ (red points), after which the increasing trend show clear transition; $\delta_Z$ (green dots) enter the regime of $\sim 1/M$ (green line) indicated by equation \ref{eqn:h layer} (a) in the vicinity of  $\delta_Z=0.5$ (green points); $\delta_R$ (blue dots) enter the regime of $\sim 1/\sqrt{M}$ (blue line) indicated by equation \ref{eqn:h layer} (b) in the vicinity of  $\delta_R=0.5$ (green points).
In that context, it is reasonable to select these demarcation curves separating different areas.

\subsection{Analytical theory for flow patterns}
This section is devoted to the comprehensive analytical study on flow patterns in inertialess regime.
We will summarize previous analytical theory or develop new ones for each pattern.

\begin{figure}
	\centerline{\includegraphics[width=\textwidth]{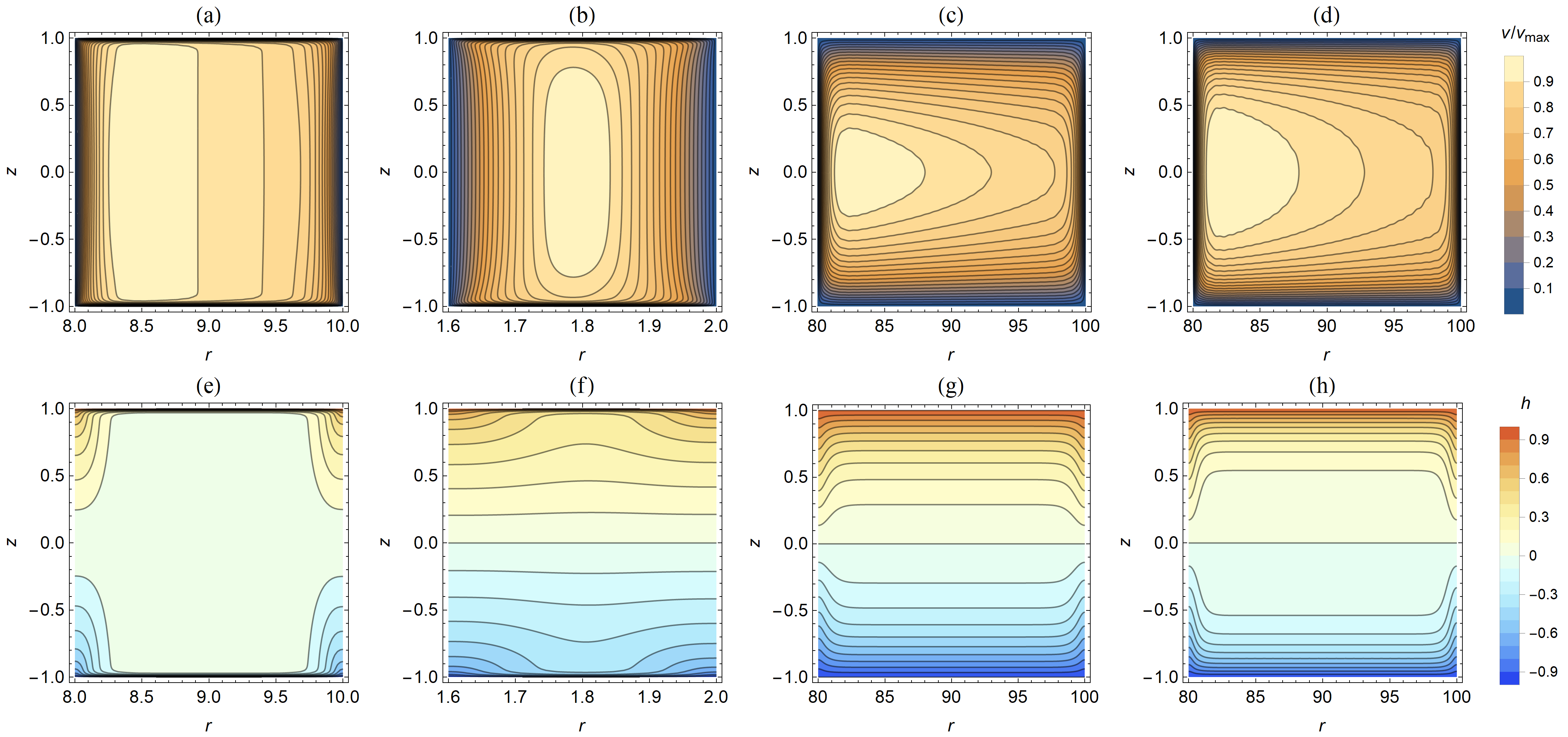}}
	\caption{Top row: contours of the modified velocity $v/v_\text{max}$; bottom row: current stream $h$ in coupled mode. 
	first column: $\eta=1, M=80$ in CRZ area; 
	second column: $\eta=5, M=80$ in CZ area;
	third column: $\eta=0.1, M=3$ in CR area;
	forth column: $\eta=1, M=3$ in CN area.}
	\label{fig:vhc}
\end{figure}

Figure \ref{fig:vhc} illustrates the velocity and current stream of 4 patterns in coupled mode in C-areas.
For these patterns, the coupling between flow field and electric field must be concerned according to equation \ref{eqn:uh-il}.
\begin{itemize}
	\item \textbf{CRZ Pattern:}
In CRZ pattern, flow profile shows the boundary layers in both R direction and Z direction.
This pattern only exists at $M\gg 1$ and $\sqrt{M}\gg\eta$ as equation \ref{eqn:h layer} indicates, in other words, above the blue and green curves figure \ref{fig:map} (a).
A representative case with $\eta=1$ and $M=80$ in CRZ area is examined. 
Contours of modified velocity $v/v_{\max}$ ($v_{\max}$ is the peak of $v$) and current stream $h$ are shown in figure \ref{fig:vhc} (a) and (d), respectively.
The Hartmann layers adjacent to end walls and Shercliff layers adjacent to side walls coexist in the velocity field.
Besides, the current stream field also see the concentration within Hartmann layers region, and great bend near Shercliff layers region.
This structure, which is solved out by our spectral method, is in good agreement with the numerical simulations of \citet{khalzov_equilibrium_2010} and \citet{zhao_instabilities_2012}.
In terms of $\bar{v}$ law of CRZ pattern, \citet{baylis_mhd_1971} employed boundary layer analysis method and obtained the analytical formula:     
\begin{equation}\label{eqn:v:CRZ}
	\bar{v}=\frac{\eta}{2} \ln{\left(\frac{1+\kappa}{1-\kappa }\right)} \left(1-\frac{1.912 \eta \kappa}{\sqrt{M} (1-\kappa ^2) \ln{\left(\frac{1+\kappa}{1-\kappa }\right)} }+O(\frac{1}{M})\right)
\end{equation}
This formula implicates that $\bar{v}$ is free of $M$ when $M$ is large, which explain the almost upright contours shown in CRZ area of figure \ref{fig:map} (a).

\item \textbf{CZ Pattern}
Above the blue curve but below green curve in figure \ref{fig:map} (a) is CZ pattern, in which $M\gg 1$ but $\sqrt{M}\sim\eta$.
A representative case with $\eta=5$ and $M=80$ is examined. 
Contours of modified velocity $v/v_{\max}$ and current stream $h$ are shown in figure \ref{fig:vhc} (b) and (f), respectively.
As expected, Hartmann layers exist in the velocity field while Shercliff layers vanish.
Figure \ref{fig:vhc} (f) shows the current stream is squeezed near the end walls, but less bent near the side walls compared to the CRZ case shown in figure \ref{fig:vhc} (e).
Using axial-force averaging with Hartmann layers present, \citet{potherat_effective_2000} created the model for radial profile of velocity with transverse magnetic field:
\begin{equation}
	\frac{\partial^2 u}{\partial r^2}-\frac{1}{r}\frac{\partial u}{\partial r}-M u+M=0
\end{equation}
which is applicable in CZ pattern \citep{khalzov_equilibrium_2010}.
The exact solution is a combination of 1-order modified Bessel functions $\mathrm{I}_1$ and $\mathrm{K}_1$:
\begin{equation}
	u(r)=1+C_1 r \mathrm{I}_1(\sqrt{M} r)+C_2 r \mathrm{K}_1(\sqrt{M} r),\label{eqn:CH}
\end{equation}
where $C_1$ and $C_2$ are constants satisfying boundary conditions. 
It leads to the analytical formula for average velocity:
\begin{equation} \label{eqn:v:CH}
	\bar{v}=1+\left(C_1 \left(\mathrm{I}_0(\sqrt{M} \frac{1+\kappa}{\kappa\eta})-\mathrm{I}_0(\sqrt{M} \frac{1-\kappa}{\kappa\eta})\right)+C_2 \left(\mathrm{K}_0(\sqrt{M} \frac{1-\kappa}{\kappa\eta})-\mathrm{K}_0(\sqrt{M} \frac{1+\kappa}{\kappa\eta})\right) \right)/{\log(\frac{1+\kappa}{1-\kappa})}
\end{equation}
\item \textbf{CR Pattern}
The case opposite to CZ pattern is called CR pattern, which is expected to occur when $\sqrt{M}\gg\eta$ but $M\sim1$.
The representative calculating case of $v/v_{\max}$ and $h$ with $\eta=0.1$ and $M=3$ is presented in figure \ref{fig:vhc} (c) and (g).
Figure \ref{fig:vhc} (c) shows that along radial axis velocity soars within Shercliff layers, while Hartmann layers are invisible.
Contrary to CZ pattern in figure \ref{fig:vhc} (f), the current stream shown in (g) see clear bend near side walls but its axial distribution is relative even. 
The analytical theory for CR pattern will be established for the first time as followings.
Beyond Shercliff layers, equations \ref{eqn:uh-il} should be transformed into:
\begin{subeqnarray}
	\frac{\partial^2u}{\partial z^2}+M\frac{\partial h}{\partial z}=0, \quad u|_{z=\pm1}=0\\
	\frac{\partial^2h}{\partial z^2}+M\frac{\partial u}{\partial z}=0, \quad h|_{z=\pm 1}=\mp1
\end{subeqnarray}
leading to solutions:
\begin{subeqnarray}
	u_c=\coth (M)-\mathrm{csch}(M) \cosh (M z)\\
	h_c=\mathrm{csch}(M) \sinh (M z).
\end{subeqnarray}
Then the average velocity is:
\begin{equation} \label{eqn:v:CR} 
	\bar{v}=\frac{M \coth (M)-1}{2 M}\eta\ln \left(\frac{\kappa +1}{1-\kappa }\right)
\end{equation}
\item \textbf{CN Pattern}
The case with $\sqrt{M}\sim\eta$ and $M\sim1$ (corresponding to the area below blue curve and green curve but above red curve) is called CN pattern, in which no boundary layer shows up but the flow is still electrically coupled.
The representative calculating case of $v/v_{\max}$ and $h$ with $\eta=1$ and $M=3$ is presented in figure \ref{fig:vhc} (d) and (h).
The contours of velocity seem to be concentric circles; the current stream is slightly bent the channel, which differ from the decoupled case of $h=z$.
1-D analytical models will fail to depict this pattern.
Nevertheless, since the high order terms in semi-analytical operation \ref{eqn:semi-series} contribute to the surge within boundary layers, we believe the truncation can produce a approximate solution to CN pattern without boundary layers.
Appendix C discusses the convergence in details.
In fact, for the case of $\eta\sim1$ and $M\sim1$, the blue point in figure\ref{fig:con} shows that the first term can constitute about more than 90\% of the sum of series. 
With the truncation, the semi-analytical solution degenerate into a purely analytical one as:
\begin{equation}
	v=\frac{\langle g_{1},1\rangle }{\langle g_{1},g_{1}\rangle} \frac{1-\frac{\cosh \left(z \sqrt{\alpha _1 \eta ^2+M^2}\right)}{\cosh \left(\sqrt{\alpha _1 \eta ^2+M^2}\right)}}{\alpha_1 \frac{\eta}{M}+\frac{M}{\eta} \frac{\tanh \left(\sqrt{\alpha_1 \eta ^2+M^2}\right)}{\sqrt{\alpha_1 \eta ^2+M^2}}} \left(\frac{\text{J}_1(\alpha_1 \eta r)}{\text{J}_1(\alpha_1\frac{1+\kappa}{\kappa})}-\frac{\text{Y}_1(\alpha_1 \eta r)}{\text{Y}_1(\alpha_1\frac{1+\kappa}{\kappa})}\right)
\end{equation}
\end{itemize}

\begin{figure}
	\centerline{\includegraphics[width=\textwidth]{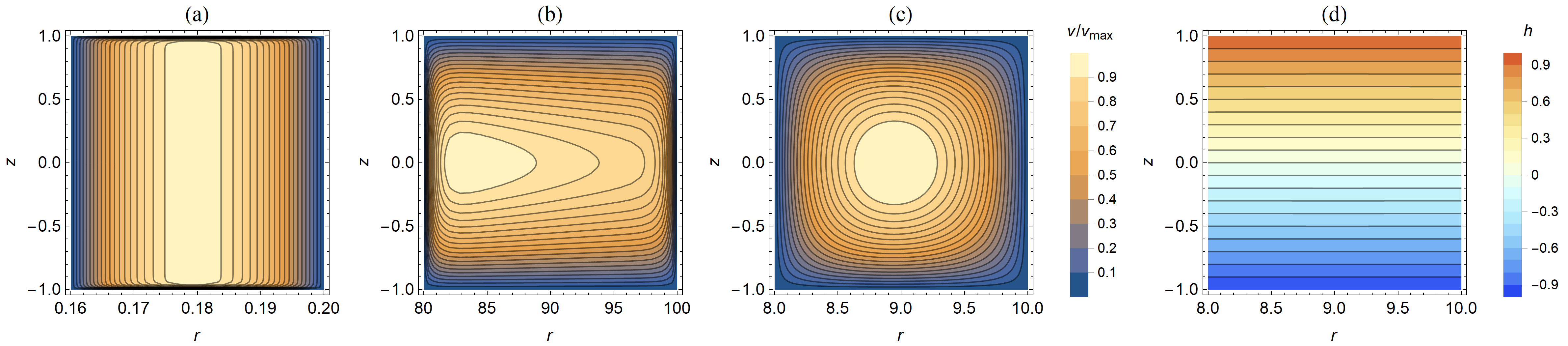}}
	\caption{Contours of modified velocity $v/\bar{v}$ of
(a): $\eta=50, M=1$ in DZ area; 
(b): $\eta=0.1, M=0.1$ in DR area;
(c): $\eta=1, M=0.1$ in DN area;
and (d) contours of current stream $h$ in decoupled regions}
	\label{fig:vhuc}
\end{figure}

Figure \ref{fig:vhuc} illustrates the velocity and current stream of 3 patterns in decoupled mode in D-areas.
Interestingly, in terms of the flow fields, DZ, DR, DN patterns are similar to CZ, CR, CN, respectively.
The reason is that their boundary layers are similar, no matter whether $\eta$ or $M$ brings about these layers.
The major difference is that the current stream in D-mode are all identical to $h=z$ as shown in figure \ref{fig:vhuc} (d), while the counterpart in C-mode show various twists or squeeze as can be seen in figure \ref{fig:vhc} (e)-(h).
For these patterns in D-areas, the electric field is decoupled from flow field and the equation \ref{eqn:uucx} shall be employed.

\begin{itemize}
	\item \textbf{DZ pattern} When $\eta\gg1$ and $M\ll\eta^2/\alpha_1$ (as indicated by equation \ref{eqn:ep2}), flow profile falls in DZ pattern.
A representative case with $\eta=50$ and $M=1$ in DZ area is examined in figure \ref{fig:vhuc} (a). 
Beyond the Z-direction boundary layers, equation \ref{eqn:uucx} is reduced to
\begin{equation}
	\eta^2\left(\frac{\partial^2 u}{\partial x^2}-\frac{1}{x}\frac{\partial u}{\partial x}\right)+M=0,\quad x=\eta r
\end{equation} 
Its analytical solution is:
\begin{equation}
	u=M \frac{\left(\kappa ^2-1\right)^2 }{\eta ^2 \kappa ^3}\ln \left(\frac{1-\kappa }{\kappa +1}\right)+r^2 \ln \left(\frac{\left(\kappa +1\right)^{\frac{(\kappa +1)^2 }{\kappa }}\left({1-\kappa }\right)^{\frac{(\kappa -1)^2 }{-\kappa }}}{(\eta\kappa)^4r^4}\right),
\end{equation}
and average velocity is
\begin{equation}\label{eqn:v:DZ}
	\bar{v}=M\frac{\kappa ^2-\left(\kappa ^2-1\right)^2 \mathrm{arctanh}^2(\kappa )}{4 \eta  \kappa ^3}\sim\frac{\kappa  M}{3 \eta }+O\left(\kappa ^3\right).
\end{equation}
This final formula implicates the slope of $\bar{v}$ contours in $M$-$\eta$ space is $-1$, as figure \ref{fig:map} (a) exactly shows.	

\item \textbf{DR pattern} When $\eta\ll1$ and $M\ll 1$ (as indicated by equation \ref{eqn:ep1}), flow profile falls in DR pattern.
A representative case with $\eta=1$ and $M=0.1$ in DR area is examined in figure \ref{fig:vhuc} (b). 
Beyond the R-direction boundary layers, equation \ref{eqn:uucx} is reduced to	
\begin{equation}
	\frac{\partial^2 u}{\partial z^2}+M=0,
\end{equation} 	
leading to a simple	solution for $u$ and $\bar{v}$
\begin{equation}
	u=M(1-z^2)/2,
\end{equation}
\begin{equation} \label{eqn:v:DR}
	\bar{v}=\frac{1}{6} \eta  M \log \left(\frac{\kappa +1}{1-\kappa }\right),
\end{equation}
which indicates the slope of $\bar{v}$ contours in $M$-$\eta$ space is $1$, as figure \ref{fig:map} (a) exactly shows.	

\item \textbf{DN pattern} When $\eta$ is moderate, there are not any boundary layers in decoupling mode regards to DN area.
A representative case with $\eta=1$ and $M=0.1$ in DN area is examined in figure \ref{fig:vhuc} (c). 
Similar to CN pattern, 2-D effect must be considered.
We solve the decoupled equation \ref{eqn:uucx} by the method of variable separating and obtain:    
\begin{equation}\label{eqn:su}
	u(r,z)=M \sum_{i=1}^{\infty}\frac{\langle u_{r,i},1\rangle}{\eta ^2 \alpha _i^2} u_{r,i}(r)\left(1-\frac{\cosh (\eta  \alpha _i  z)}{\cosh (\eta  \alpha _i)}\right)
\end{equation}
With this series solution, we can figure out the average velocity:
\begin{equation}
	\bar{v}=M \sum_{i=1}^{\infty}\frac{\langle u_{r,i},1\rangle^2}{2\alpha _i^2}\left(1-\frac{\tanh \left(\eta  \alpha _i\right)}{\eta  \alpha _i}\right)\frac{1}{\eta }
\end{equation}
On the same ground with CN pattern, the first-order approach can be taken as:
\begin{equation}\label{eqn:v:DN}
	\bar{v}\sim\left(1-\frac{\tanh \left(\eta  \alpha _1\right)}{\eta  \alpha _1}\right)\frac{M}{\eta }
\end{equation}
\end{itemize}


\subsection{Applicability of semi-analytical solution}
\begin{figure}
	\centerline{\includegraphics[width=0.6\textwidth]{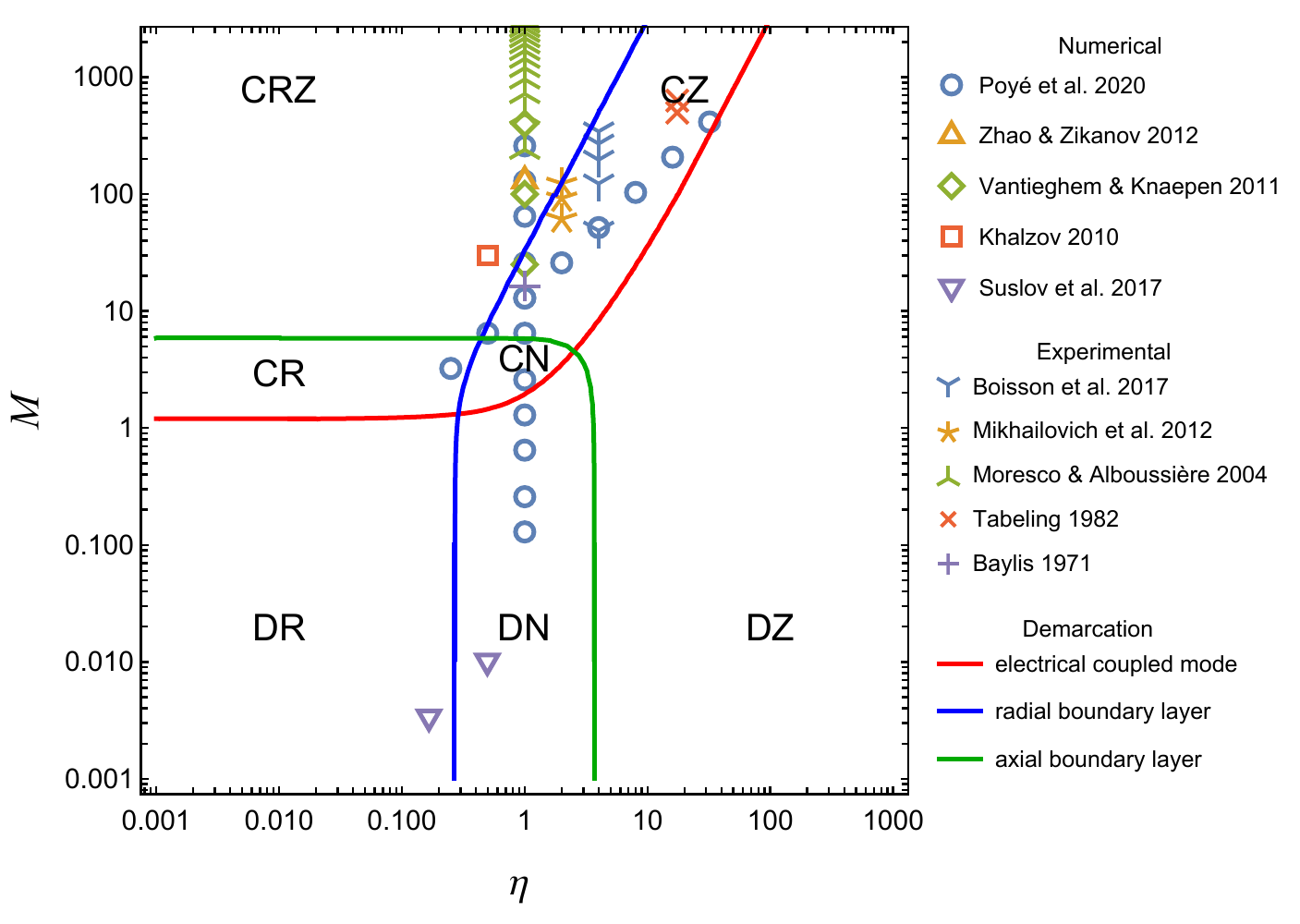}}
	\caption{Flow patterns in inertial regime. The hollow markers, asterisks, coloured areas represent previous numerical, experimental and analytical studies respectively.}
	\label{fig:summary}
\end{figure}

Figure \ref{fig:summary} summarizes this current work together with previous investigations.
The hollow markers and asterisks are previous numerical and experimental studies, respectively. 
It is noticed that previous work focused mainly on the CRZ and CZ patterns in a square cross-section with $\eta=1$.
Indeed, previous simulations were in principle also able to be applied for any single points in $M$-$\eta$ space.
The semi-analytical solution proposed in this paper is supposed to recover previous data, and more importantly, cover all areas in the whole parameter space once and for all. 

\begin{figure}
	\centerline{\includegraphics[width=0.9\textwidth]{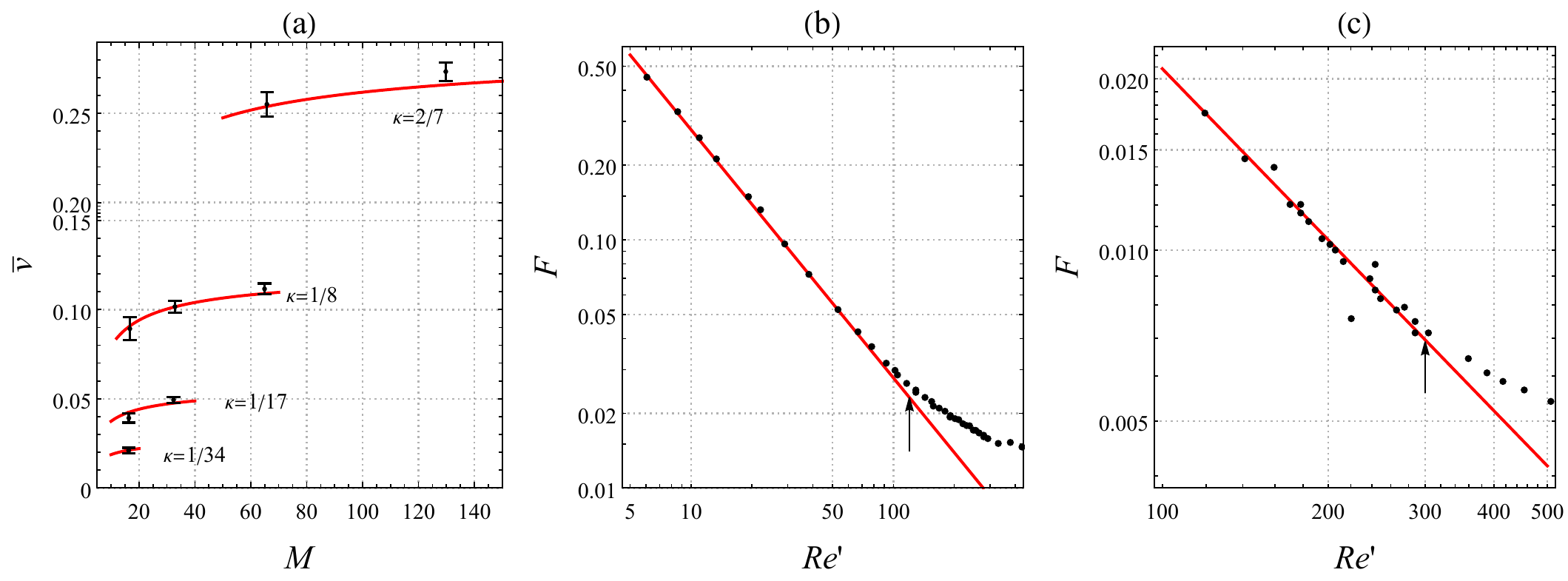}}
	\caption{comparison between spectral solution (red curve) and experiment data (black points).
		(a) variations of $\bar{v}$ with Hartmann number $M$ at $\eta=1$ and $\kappa=1/34, 1/17, 1/8, 2/7$ compared with \citep{baylis_mhd_1971}. (b) Friction factor $F$ vs $Re'$ compared with \citep{baylis_experiments_1971}. (c) Friction factor $F$ vs $Re'$ compared with \citep{moresco_experimental_2004}. The arrows in (b) and (c) indicate transition points from laminar flow to turbulence.}
	\label{fig:cexp}
\end{figure}
\begin{figure}
	\centerline{\includegraphics[width=0.8\textwidth]{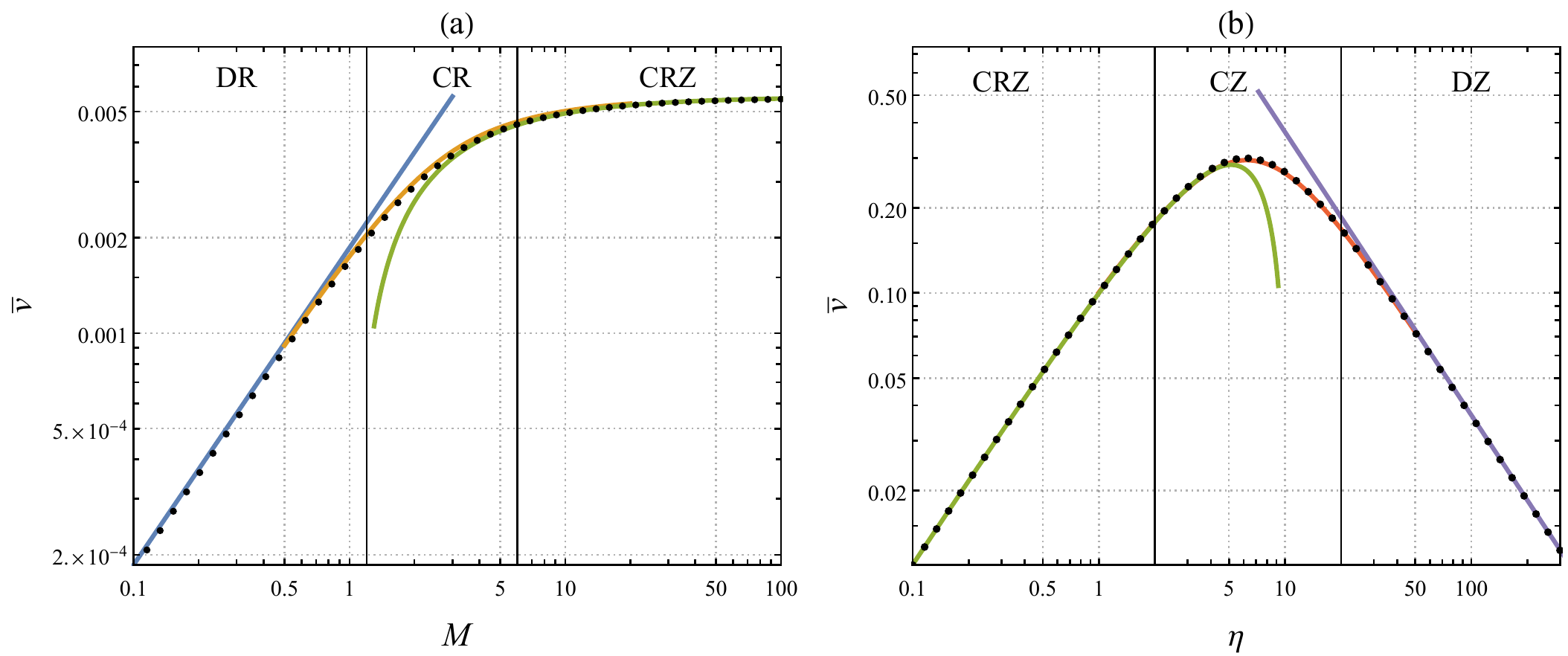}}
	\caption{Comparison between spectral solution (black dots) and analytical theory (coloured lines: equation \ref{eqn:v:DR} (blue), equation \ref{eqn:v:CR}  (yellow), equation \ref{eqn:v:CRZ} (green), equation \ref{eqn:v:CH} (red), equation \ref{eqn:v:DZ} (purpler)). (a) Variations of $\bar{v}$ with Hartmann number $M$ across DR, CS,and CRZ regions when $\eta=0.05$. (b) Variations of $\bar{v}$ with geometric ratio $\eta$ across CRZ, CH,and DZ regions when $M=100$.}
	\label{fig:cana}
\end{figure}
The present semi-analytical solution is verified by the comparison with experiment data and analytical theory. 
On the one hand, there is amount of experiment data in CRZ region with great $M$ and square section ($\eta=1$).
The comparison between spectral solution and experiment in CRZ region is shown in figure \ref{fig:cexp}.
Figure \ref{fig:cexp}(a) compared the variations of $\bar{v}$ with Hartmann number $M$ and $\kappa$ obtained by spectral solution and \citet{baylis_mhd_1971} experiment. 
The growing tendency of $\bar{v}$ with increasing $M$ and $\kappa$ coincides with experiment considering error bars.  
Another important parameter, friction factor which is defied as  
\begin{equation}
	F=\frac{I_0B_0}{2\pi \rho\bar{R} (V_0\bar{v})^2 }=\frac{M}{Re}\frac{2\kappa\eta}{\bar{v}},
\end{equation}
is concerned by the experiment of \citet{baylis_experiments_1971} and \citet{moresco_experimental_2004}. 
The relationship of $F$ vs. $Re'$, where $Re'=Re/M$, is commonly used to detect transition from laminar flow to turbulence.
Figure \ref{fig:cexp}(b) and (c) compared this relationship calculated by the semi-analytic spectral solution and that measured by experiments.
Note that the semi-analytic algorithm is based on laminar theory, while these experiments detected the transition from laminar flow to turbulence in the inertialess regime.  
The clear deviation between red curve and black points is therefore seen in figure \ref{fig:cexp}(b) and (c), in spite of the good agreement before transition point labelled by the arrows.  

On the other hand, the spectral solution can be verified by the analytical theory \ref{eqn:v:CRZ} for CRZ pattern, equation \ref{eqn:v:CH} for CZ pattern, equation \ref{eqn:v:CR}  for CR  pattern, equation \ref{eqn:v:DR} for DR pattern, and equation \ref{eqn:v:DZ} for DZ pattern.
Figure \ref{fig:cana}(a) shows $\bar{v}$ vs. $M$ with $\eta=0.05$.
It is noticed that scale laws of $\bar{v}$ vary with Hartmann number $M$. 
To be specific, when $M$ is small (DR pattern), a clear linear relationship between $\lg\bar{v}$ and $\lg M$ can be seen as depicted by the blue line according to analytical approach \ref{eqn:v:DR}. 
When $M$ is large (CRZ pattern), the scale of $\bar{v}$ is insensitive to the increase in $M$ as depicted by the green curve according to analytical approach \ref{eqn:v:CRZ}. 
However, both equation \ref{eqn:v:CRZ} and \ref{eqn:v:DR} fail in the transition zone (CR  pattern), where the preciser analytical theory \ref{eqn:v:CR}  (yellow curve) works.
Figure \ref{fig:cana}(b) shows $\bar{v}$ vs. $\eta$ with $M=100$.
The scale of $\bar{v}$ has a peak value in the proximity of $\eta=6$, before which there is the linear increase in $\lg\bar{v}$ as controlled by CRZ theory \ref{eqn:v:CRZ}.
Inversely, if $\eta$ rises into DZ region, $\lg\bar{v}$ will see a linear decline as depicted by the purple line of analytical theory \ref{eqn:v:DZ}.
Similar to the case shown in (a), both equation \ref{eqn:v:CRZ} and \ref{eqn:v:DR} fail in the transition zone (CZ pattern), where another preciser analytical theory \ref{eqn:v:CH} (red curve) works.

It is an important identification that our semi-analytic results (black plus signs) coincide exactly with the specific analytical theory across all the six patterns with smooth transition.
\section{Results: Inertial Regime}
\label{sec:inertial}
In inertial regime, the secondary flow and its effect on primary flow must be concerned.
According to the perturbation expansion \ref{eqn:pertubation} described in section \ref{sec:method}, we express $u$ and $w$ as the series of $Re$.
Appendix D is devoted to the estimation of convergence of the series.
Interestingly, the mathematical analysis tells that the convergence rests with $Re'=Re/M$ instead of $Re$.
This $Re'$ is exactly the variable indicating the transition from laminar to turbulence \citep{moresco_experimental_2004}.
The estimation we draw in Appendix D shows the convergence domain is about $Re'<\kappa^{-2.5}$ ($=243$ in the case with $\kappa=1/9$).
This convergence domain is slightly less than the transition point, at $Re'\approx380$, in the experiment of \citet{moresco_experimental_2004}.
Because the current research focuses on the laminar case, the convergence domain
\begin{equation}
	 Re'<\kappa^{-2.5}
\end{equation}
is a conservative estimate for the applicability of the proposed perturbation method.
Moreover, it is noticed that the even orders of $Re$ vanish in the series for $u$ while odd orders vanish in the series for $w$ equations\ref{eqn:pertubation}:
\begin{eqnarray}
	u=u_0+u_2Re^2+u_4Re^4+...\\
	w=w_1Re+w_3Re^3+...
\end{eqnarray}
The scale analysis shows that:
\begin{equation}
	\frac{u_4Re^4}{u_2Re^2}\sim \frac{w_3Re^3}{w_1Re}\sim\kappa^5Re'^2
\end{equation}
In the case of $\kappa^5Re'^2\ll1$, which can be reached easily at small $\kappa$, truncating the series is reasonable for examining the effect of $Re$:
\begin{eqnarray}\label{eqn:truncating re}
	u=u_0+u_2Re^2\\
	w=w_1Re
\end{eqnarray}
The following results are calculated with fixed geometric parameter $\eta=1$ and $\kappa=1/9$, which are the same as the investigation of \citet{moresco_experimental_2004,zhao_instabilities_2012,vantieghem_numerical_2011}.

Again, we must \textbf{emphasize} that there is no restriction on the magnitudes of $Re$ nor $M$ per se. 
As long as $Re'$ is not too great, the flow remains laminar\citep{moresco_experimental_2004}, and the proposed perturbation solutions always converge.

\subsection{evolution of secondary flow}
\begin{figure}
	\centerline{\includegraphics[width=\textwidth]{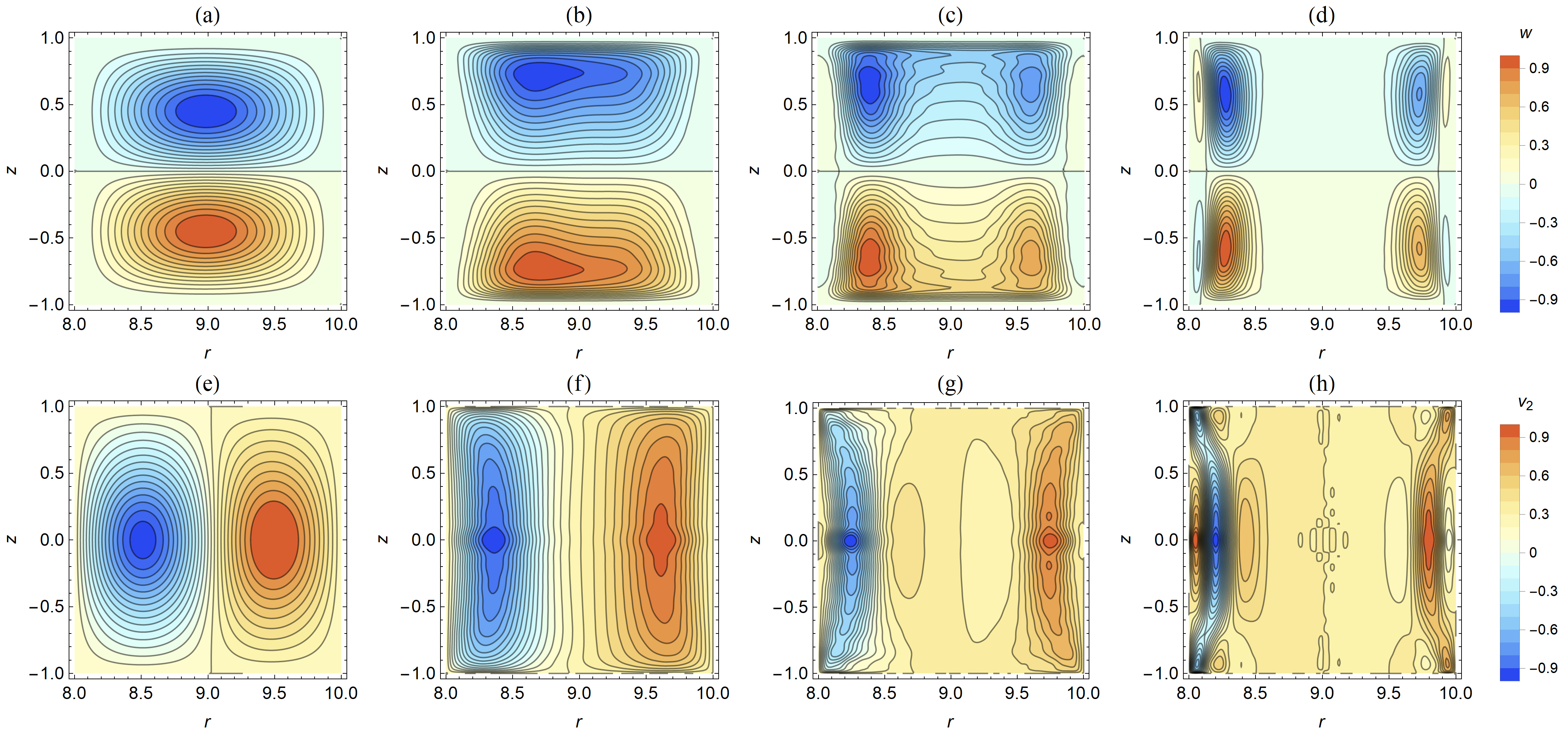}}
	\caption{Top row: stream function $w_1$ for secondary flow; bottom row: perturbation velocity $v_2$ driven by inertial effect. First column: $M=1$; second column: $M=20$; third column: $M=50$; forth column: $M=100$;}
	\label{fig:wv}
\end{figure}

Figure \ref{fig:wv} (a)-(h) list the stream function $w_1$ of secondary flow on the top row, and velocity perturbation $v_2=u_2/r$ due to inertial effect on the bottom row with incremental $M$ from left to right.
Note that $v_2$ and $w_1$ are not controlled by $Re$ according to equation \ref{eqn:truncating re}.
This subsection is devoted to the structure of flow field and the magnitude will be discussed in the next subsection.
$w$ and $v_2$ in figure \ref{fig:wv} is normalized by the maximum and thus range from $-1$ to $1$.
It can be seen that the flow profiles vary with $M$ and there are four different patterns, which are named oval, trapezoid, cracking, and separation patterns based on shape of secondary vortexes.
\begin{itemize}
	\item \textbf{Oval pattern} Figure \ref{fig:wv}(a) and (e) depict the inertial case with $M\sim1$ corresponding to DN pattern in inertialess regime.
	Figure \ref{fig:wv}(a) shows a pair of oval antisymmetry secondary vortexes.
	Then the two large secondary vortexes modify the profile of primary flow via Ekman pump effect, enhancing azimuthal velocity on the outer side and weaken that on the inner, as shown in figure \ref{fig:wv}(e).
	\item \textbf{Trapezoid pattern} Figure \ref{fig:wv}(b) and (f) depict the inertial case with $M\sim10$ corresponding to CZ pattern in inertialess regime.
	For the primary flow, Hartmann layers emerge near the insulation end walls.
	The rapid change in primary flow near Hartmann layers leads to the dramatic variation in $v_r$ of secondary flow near the plate ends, as indicated by high density of contours in figure \ref{fig:wv}(b).
	The antisymmetry secondary vortexes transformed into trapezoid, which coincides with the simulations of \citet{vantieghem_numerical_2011}.
	The perturbation velocity $v_2$ stretches axially and has radial one-dimensional profile in the core region as can be seen in figure figure \ref{fig:wv}(f).
	\item \textbf{Cracking pattern} As Hartmann number goes up to $M\sim50$, the trapezoid vortexes have begun to crack and given birth to a second small peak near the outer side as shown in figure \ref{fig:wv}(c). 
	This cracking phenomenon agree with the simulations of \citet{zhao_instabilities_2012}.
	Note that reverse vortexes adjacent to conductive torus walls also emerge along with the main vortexes cracking.
	The reverse vortexes can be identified by the separated zero-contours near radial boundaries and we will discuss their critical emergence later.
	And in this pattern, figure \ref{fig:wv}(g) shows that the perturbation velocity $v_2$ is crushed radially resulting in dramatic perturbation near the torus surfaces but marginal effect in the middle region.
	Figure \ref{fig:wv}(g) can explain the results of \citet{zhao_instabilities_2012}: inertial effect thicken inner Shercliff layer while thin outer Shercliff layer.
	\item \textbf{Separation pattern} When Hartmann number is large ($M\sim100$), the main vortexes are broken into two separated parts and reverse vortexes are fully developed as shown in figure \ref{fig:wv}(d).
	The anomalous reverse vortexes conform the theoretical prediction of \citet{tabeling_magnetohydrodynamic_1981}. 
	It can be seen in figure \ref{fig:wv}(h) that $v_2$ has a drastic change from positive to negative within Shercliff layers.
	In this context, inertial effect can strengthen primary velocity near the inner side, which is opposite to other patterns stated before.
	This anomaly can be understood through the reverse vortexes.
	According to equation \ref{eqn:iu} and the perturbation expansion in section \ref{sec:method}, the equation for $v_2$ can be expressed as:
	\begin{equation}
		0=\Delta^*({r}{v_2})+M\frac{\partial h_2}{\partial z}+\frac{1}{r}\left(\frac{\partial u_0}{\partial r}\frac{\partial w_1}{\partial z}-\frac{\partial u_0}{\partial z}\frac{\partial w_1}{\partial r}\right)
	\end{equation}
	Note that $\partial_r u$ is great within Shercliff layers when $M$ is large. 
	And the reverse vortexes of secondary flow raise the dramatic signed shift of $\partial_z w$ along the middle axis $z=0$ near Shercliff layers.
	Hence, there is the reversal sign in source item $\partial_r u \partial_z w$, which leads to the anomalous inertial perturbation on primary flow.
\end{itemize}

\subsection{scale formula of suppression of inertial effect}
\begin{figure}
	\centerline{\includegraphics[width=0.7\textwidth]{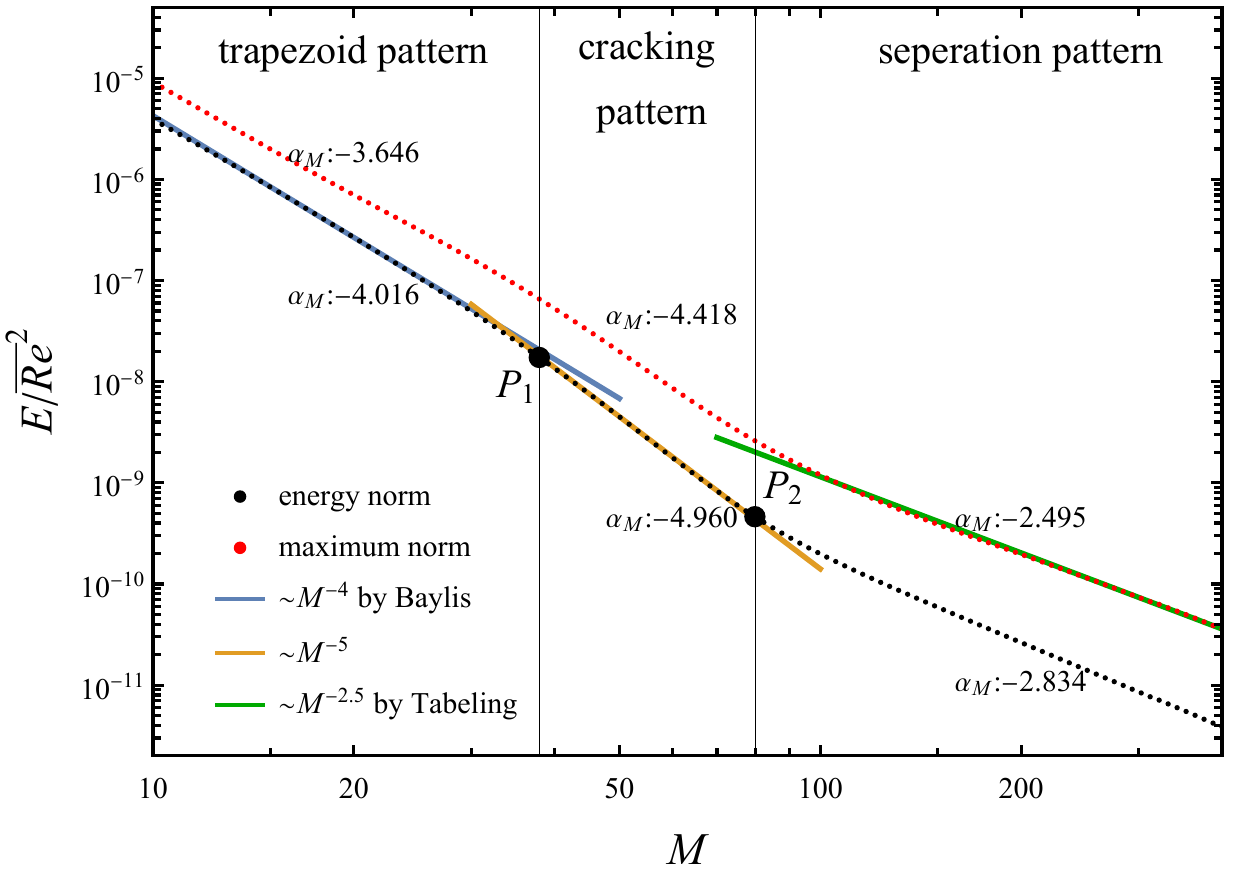}}
	\caption{Scale of inertial perturbation on primary velocity. Black and red points are spectral results according to equation \ref{eqn:E energy} and equation \ref{eqn:E max}, respectively. Blue line are $\alpha_M={-4}$ theory of \citet{baylis_mhd_1971}; green line is $\alpha_M={-2.5}$ theory of \citet{tabeling_magnetohydrodynamic_1981}; yellow line with $\alpha_M={-5}$. All the label $\alpha_M:XXX$ comes from local linear fitting for the energy norm or maximum norm.}
	\label{fig:magn}
\end{figure}

We define $E$ to estimate the scale of inertial effect:
\begin{equation}\label{eqn:E energy}
	E=\bar{Re}^2\frac{||v_2||}{||v_0||},
\end{equation}
where $\bar{Re}=v_0 Re $ is spatially average Reynolds number, $||f||$ is the energy norm defined in equation \ref{eqn:energy norm}.
The norm operator is not one and only. 
The maximum norm can also be used:
\begin{equation}\label{eqn:E max}
	E=\bar{Re}^2\frac{\max\{v_2\}}{\max\{v_0\}},
\end{equation}
There is no final conclusion about which norm operator is better to estimate the scale of inertial effect.
The contribution of Reynolds number to inertial effect is clear as $E\sim\bar{Re}^2$, while the effect of Hartman number is latent.
Based on scale analysis of Hartmann layers, the theory of \citet{baylis_mhd_1971} declare:
\begin{equation}\label{eqn:magn:b}
	E\sim{\bar{Re}^2}/{M^4}.
\end{equation}
By contrast, based on scale analysis of Shercliff layers, the theory of \citet{tabeling_magnetohydrodynamic_1981} declare:
\begin{equation}\label{eqn:magn:t}
	E\sim{\bar{Re}^2}/{M^{2.5}}
\end{equation}
In order to examine the index of Hartmann number
\begin{equation}
	\alpha_M=\frac{\partial\lg E/\bar{Re}^2}{\partial\lg M},
\end{equation}
figure \ref{fig:magn} presents $E/\bar{Re}^2$ vs. $M$ in logarithmic plot.
120 cases with different $M$ are calculated in this inertial regime.
The black points are calculated by equation \ref{eqn:E energy} based on $v_2$ and $v_0$ results.
Blue and green lines refer to energy norm \ref{eqn:magn:b} and maximum norm \ref{eqn:magn:t}, respectively.
It can be seen that the increasing $M$ suppresses the magnitude of inertial perturbation.
The spectral results agree with $M^{-4}$ theory when $M<40$ (before $P_1$), but drop faster than $M^{-4}$ when $40<M<80$ (between $P_1$ and $P_2$).
The linear fitting of spectral results gives the relationship of $\sim M^{-5}$.
When $M>80$, the suppression effect of $M$ gets weaker and $E$ declines slower than $M^{-4}$.
There is still a little deviation from $M^{-2.5}$ theory of \citet{tabeling_magnetohydrodynamic_1981}.
However, if the maximum norm equation \ref{eqn:E max} is taken, we can confirm the $M^{-2.5}$ theory as shown by red points.
To be precise, each section of the data is labelled with $\alpha_M$ produced by linear fitting.
Overall, maximum norm indicates slower suppression laws with smaller $|\alpha_M|$ for each section than those of the energy norm.
The reason is presented in the following subsection \ref{sec:relation}.
\subsection{relationship between scale formula and secondary flow}\label{sec:relation}
\begin{figure}
	\centerline{\includegraphics[width=0.8\textwidth]{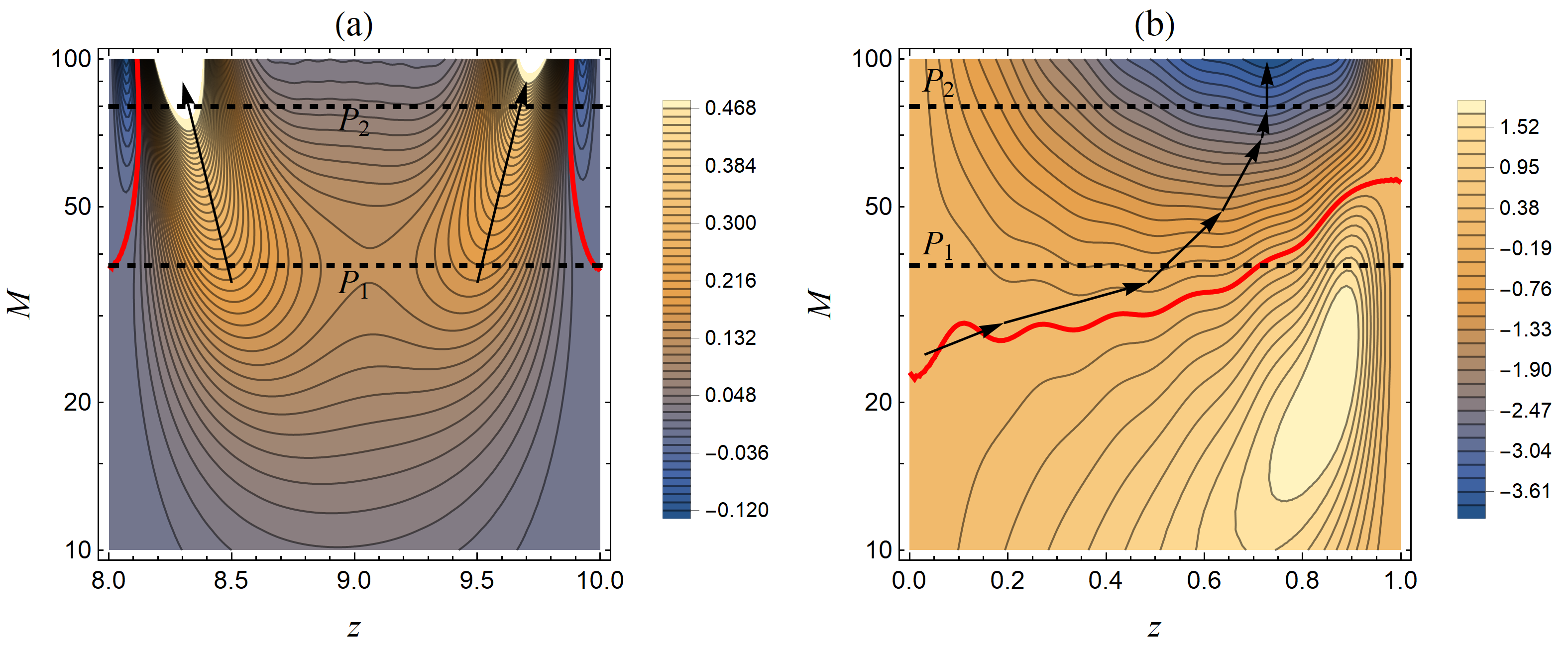}}
	\caption{(a) radial profile of stream function $w$ of secondary flow with varying $M$ at $z=0$ . (b) axial profile of azimuthal vorticity $\mathrm{rot}_\theta \mathbf{v}$ with varying $M$ on the inner torus wall (at $r=r_1$). Red curves are zero-contour.}
	\label{fig:vortex}
\end{figure}

The shift of the scale formula can be understood through the varied patterns of secondary flow.
The analysis of \citet{baylis_mhd_1971} takes account into $u_r$ but neglects $u_z$, and finally deduces $E\sim M^{-4}$.
Before the critical point $P_1$, the secondary flow is trapezoid pattern, in which $u_r$ near Hartmann layers is predominant, as shown in figure \ref{fig:wv} (e).
In this context $M^{-4}$ theory is accurate in trapezoid pattern.
To detect the exact shift point $P_1$ when vortexes begin cracking, we define radial profile of secondary flow:
\begin{equation}
	w_r=M^4\int_{0}^{1}w_1\mathrm{d} z.
\end{equation}
The continuos variations of $w_r$ with Hartmann number are shown in figure \ref{fig:vortex} (a).
We can see that when $M\approx30\sim40$, the single peak of $w$ crack into two peaks.
This is also when $M^{-4}$ theory becomes invalid.
As $M$ rises, the two peaks move away from each other.
Meanwhile, $w$ become lower in the middle region, leading to the low $v_2$ in middle region shown in figure \ref{fig:wv}(f).
This low value region expands as $M$ increases.
Hence, the suppression of $M$ on inertial perturbation become faster as shown in $P_1$-$P_2$ part of figure \ref{fig:magn}.
However, when $M$ becomes larger than 80 (after $P_2$), the expansion of low-value region stops while the reverse vortexes get larger.
This is the separation pattern as stated before.
The reverse vortexes bring in great $u_z$ near Shercliff layer.
The analysis of \citet{tabeling_magnetohydrodynamic_1981} takes account into $u_z$ then deduces $E\sim M^{-2.5}$.
In this context, $M^{-2.5}$ theory is accurate in separation pattern.
Moreover, since $M^{-2.5}$ theory only concerns the perturbation near Shercliff layer, the maximum norm (red points in figure \ref{fig:magn}) gives closer results than spatial-average energy norm.

To depict how the reverse vortexes develop, the azimuthal vorticity, 
\begin{equation}
	\mathrm{rot}_\theta \mathbf{v}=\frac{\partial v_r}{\partial z}-\frac{\partial v_z}{\partial r},
\end{equation}
are calculated with different $M$ on the inner torus wall.
Figure \ref{fig:vortex} (b) shows the results of $M^2\mathrm{rot}_\theta \mathbf{v}$.
The coefficient $M^2$ aims at clearer contours to offset the decrease in magnitude of vorticity.
The red curves are zero-value contours, below which no reverse vortex exists.
It is noticed that the reverse vortexes partly occur near $z=0$ when $M\approx 20$.
Then as $M$ increases, the reverse region expands to the whole range of $z=0\sim1$.
$M=38$ is an important point when the sign of average vorticity shifts.
This critical point is also very close to $P_1$ where trapezoid pattern shifts into cracking pattern.
The peak of reverse vorticity moves outwards as arrows show.
When $M>80$, the place of peak value is static, which marks the entry into separation pattern and $\sim M^{-2.5}$ theory.

\section{Conclusions}
In terms of method, we have proposed a cost-effective semi-analytical algorithm for the MHD flow in annular channel.
Two sub-schemes are involved: the Spectral-Galerkin scheme for primary flow and the perturbation method allowing for the secondary flow and inertial effect.
The former has an unconditional convergence, while the convergence domain of the latter is proved to be $Re'=Re/M<\kappa^{-2.5}$.
The semi-analytical solution is validated by the comparison with experiment and simulation.
The fast computation makes it practicable to perform a large number of simulations with continuously changing operation conditions, which contribute to the exploration of flow pattern and scale laws. 

In inertialess regime, a posteriori analysis of flow patterns has been conducted. 
First, the map of average velocity based on 40,000 cases in $\eta$-$M$ space is the main result.
The phenomenological classification of flow patterns become clear at a glance in the map.
Second, the electrically coupling demarcation of $\eta$-$M$ space is drawn.
This curve indicates flow field and electric field can be decoupled when $M\ll1$ or $M/\eta^2\ll 1$, while the curvature $\kappa$ has marginal effect on the electrically coupled modes.
Third, seven distinct flow patterns are identified.
When $\eta\ll 1$ and $M$ is not large, only Shercliff layers exist and this pattern is examined for the first time.
The proposed semi-analytic spectral solution recovers all patterns throughout the whole $\eta$-$M$ space.

In inertial regime, we examine the relationship between suppression law on inertial effect and the condition of vortexes in secondary flow.
120 cases with different Hartmann numbers ranging form 1 to 400 are calculated to explore the suppression law.
It is identified that as $M$ increase, the vortices of secondary flow undergo four different conditions:
(1) a pair of oval toroidal vortices, (2) a pair of trapezoid vortices, (3) two pairs of vortices, (4) two pairs of main large vortices with two pairs of thin anomalous reverse vortices.
In the first two stage of oval and trapezoid vortices, the proposed semi-analytic spectral solution coincides with the $Re^2/M^{4}$ suppression theory of \citet{baylis_mhd_1971}.
However, as $M$ rise up to $40$, the pair of trapezoid vortices begin to crack into two pairs.
Semi-analytic solutions see faster drop in inertial effect as $Re^2/M^{5}$.
After that, when $M>80$, the anomalous reverse vortices are fully developed near Shercliff layers resulting in the slower suppression mode of $Re^2/M^{2.5}$, which coincides with the theory of \citet{tabeling_magnetohydrodynamic_1981}.

\section*{Appendix A: basis function}\label{sec:AA}
Because of the radial boundary coordinates $r=(1\pm\kappa)/{\kappa\eta}$, it is convenient to set $x=\eta r$ and $g_n(x)=u_{r,n}(r)$ and $s_n(x)=h_{r,n}(r)$.
Substituting this transformation into equation\ref{eqn:uh-u}, we can get two Sturm-Liouville system for the radial function $g$ as:
\begin{equation}
	\frac{\text{d} }{\text{d} x}\left(\frac{1}{x}\frac{\text{d} g_n}{\text{d} x}\right)=-\alpha_n^2\frac{1}{x}g_n, \quad \left. g_n\right|_{x=\frac{1\pm\kappa}{\kappa}}=0
\end{equation}
with the eigenvalues $\alpha_n$ and the weight function $1/x$.
It leads to the orthogonal function space $g_n, n=1,2,3...$ and the normalized solution:
\begin{equation}
	g_n(x;\kappa,\alpha_n)=\frac{x\text{J}_1(\alpha_nx)}{\text{J}_1(\alpha_n\frac{1+\kappa}{\kappa})}-\frac{x\text{Y}_1(\alpha_nx)}{\text{Y}_1(\alpha_n\frac{1+\kappa}{\kappa})},
\end{equation}
where $\text{J}_1$ is the 1-order Bessel function and $\text{Y}_1$ is the 1-order Neumann function. 
The eigenvalues $\alpha_n$ are determined by the boundary condition:
\begin{equation}
	g_n(\frac{1-\kappa}{\kappa};\kappa,\alpha_n)=0.
\end{equation}
When $n\gg1$, there is the approximate formula:
\begin{equation}
	\alpha_n\approx\frac{n\pi}{2}
\end{equation}

Similarly, for $s_n(x)$ we have:
\begin{equation}
	\frac{\text{d} }{\text{d} x}\left(\frac{1}{x}\frac{\text{d} s_n}{\text{d} x}\right)=-\beta_n^2\frac{1}{x}s_n, \quad
	\left. \frac{\text{d}s_n}{\text{d} x}\right|_{x=\frac{1\pm\kappa}{\kappa}}=0
\end{equation}
with the eigenvalues $\beta_n$ and the weight function $1/x$. It leads to the orthogonal function space $s_n, n=1,2,3...$:
\begin{equation}
	s_n(x;\kappa,\beta_n)=\frac{x\text{J}_1(\beta_nx)}{\text{J}_0(\beta_n\frac{1+\kappa}{\kappa})}-\frac{x\text{Y}_1(\beta_nx)}{\text{Y}_0(\beta_n\frac{1+\kappa}{\kappa})},
\end{equation}
The eigenvalues $\beta_n$ are determined by the boundary condition:
\begin{equation}
	\partial_xs_n(\frac{1-\kappa}{\kappa};\kappa,\beta_n)=0.
\end{equation}
Note that $\beta_1=0$ and $s_1(x;\kappa,0)\rightarrow1$.
When $n\gg1$, there is the approximate formula:
\begin{equation}
	\beta_n\approx\frac{(n-1)\pi}{2}
\end{equation}

Then we define the integral operator:
\begin{equation}
	\langle f_1,f_2\rangle\equiv\int_{r_1}^{r_2}f_1f_2/r\text{d} r\equiv\int_{\frac{1-\kappa}{\kappa}}^{\frac{1+\kappa}{\kappa}}f_1f_2/x\text{d} x
\end{equation}
The orthogonal law of Sturm-Liouville theory states:
\begin{equation}
	\text{if $i\neq j$, }\quad \langle g_i,g_j\rangle=\langle s_i,s_j\rangle=0.
\end{equation}
After resetting the normalized $u_{r,n}=g_n(x)/\sqrt{\langle g_n,g_n\rangle}$ and $h_{r,n}=s_n(x)/\sqrt{\langle s_n,s_n\rangle}$ we obtain the equation \ref{eqn:spectral-Bessel}

\section*{Appendix B: weak solution}\label{sec:AB}
Taking equation\ref{eqn:axial h} into the derivative of equation\ref{eqn:axial u} we obtain:
\begin{equation}\label{eqn:high order h}
	\langle u_{r,i},h_{r,j}\rangle h''''_{z,j}- \left((M^2 + \eta^2\alpha_i^2)\langle u_{r,i},h_{r,j}\rangle+\langle u_{r,i},h_{r,j}\rangle \eta^2\beta_j^2\right) h''_{z,j} +\eta^2\alpha_i^2\langle u_{r,i},h_{r,j}\rangle\eta^2\beta_j^2h_{z,j}=\langle S_w,u_{r,i}\rangle
\end{equation}
If we define the matrix as:
\begin{equation}
	\boldsymbol{u}_z=
	\begin{bmatrix}
		u_{z,1}\\
		\vdots\\
		u_{z,m}
	\end{bmatrix},
	\boldsymbol{h}_{z}=
	\begin{bmatrix}
		h_{z,1}\\
		\vdots\\
		h_{z,m}
	\end{bmatrix},
	\boldsymbol{\alpha}=
	\begin{bmatrix}
		\alpha_1^2& & \\
		&\ddots& \\
		& & \alpha_m^2\\ 
	\end{bmatrix}, 
	\boldsymbol{\beta}=
	\begin{bmatrix}
		\beta_1^2& & \\
		&\ddots& \\
		& & \beta_m^2\\ 
	\end{bmatrix}
\end{equation}
and
\begin{equation}
	\boldsymbol{\tau}=
	\begin{bmatrix}
		\tau_{i,j}\rightarrow\langle u_{r,i},h_{r,j}\rangle
	\end{bmatrix},\quad
	\boldsymbol{\gamma}=
\begin{bmatrix}
	\gamma_{i,j}\rightarrow\langle u_{r,i},h_{r,j}\rangle^{-1}\langle S_w,u_{r,i}\rangle
\end{bmatrix},\quad
	1\le i,j\le m
\end{equation}
\begin{equation}
	\boldsymbol{B}=
	\begin{bmatrix}
		0&\boldsymbol{E}\\
		-\eta^4(\boldsymbol{\tau}^{-1}\boldsymbol{\alpha}\boldsymbol{\tau})\boldsymbol{\beta}&\eta^2\boldsymbol{\beta}+\eta^2\boldsymbol{\tau}^{-1}\boldsymbol{\alpha}\boldsymbol{\tau}+M^2\boldsymbol{E}
	\end{bmatrix}
\end{equation}
where $\boldsymbol{E}$ is the $m$-order identity matrix and $\boldsymbol{B}$ is the $2m$-order square matrix. Then equation \ref{eqn:high order h} becomes 
\begin{equation}
	\frac{\text{d}^2}{\text{d} z^2}
	\begin{bmatrix}
		\boldsymbol{h}_{z}\\
		\boldsymbol{h}''_{z}
	\end{bmatrix}
	=\boldsymbol{B}
	\begin{bmatrix}
		\boldsymbol{h}_{z}\\
		\boldsymbol{h}''_{z}
	\end{bmatrix}+	
\begin{bmatrix}
	\boldsymbol{0}\\
	\boldsymbol{\gamma}
\end{bmatrix}
\end{equation}
To solve this differential equations set, we must calculate the eigenvalues $\chi$ and eigenvectors $\boldsymbol{Q}$ of $\boldsymbol{B}$:
\begin{equation}
	\boldsymbol{Q}^{-1}\boldsymbol{B}\boldsymbol{Q}=
	\underset{1\le i\le 2m}{\mathrm{Diag}}
	\begin{bmatrix}
		\chi_i
	\end{bmatrix}
\end{equation}
The solution is:
\begin{equation}
	\begin{bmatrix}
		\boldsymbol{h}_{z}\\
		\boldsymbol{h}''_{z}
	\end{bmatrix}
	=
	\boldsymbol{Q}
	\underset{1\le i\le m}{\mathrm{Diag}}
	\begin{bmatrix}
		\frac{\sinh(\sqrt{\chi_i}z)}{\sinh(\sqrt{\chi_i})}
	\end{bmatrix}
	\boldsymbol{C}
\end{equation}
Given $h$, the solution to $u$ is:
\begin{equation}
	\boldsymbol{u}_z
	=\frac{1}{\eta^2M}\boldsymbol{\alpha}^{-1}\boldsymbol{\tau}
	\begin{bmatrix}
		M^2\boldsymbol{E}+\eta^2 \boldsymbol{\beta};&-\boldsymbol{E}
	\end{bmatrix}
	\boldsymbol{Q}
	\underset{1\le i\le m}{\mathrm{Diag}}
	\begin{bmatrix}
		\sqrt{\chi_i}\frac{\cosh(\sqrt{\chi_i}z)}{\sinh(\sqrt{\chi_i})}
	\end{bmatrix}
	\boldsymbol{C}
\end{equation}
where $\boldsymbol{C}$ is a constant array that shall be determined by the boundary condition\ref{eqn:BC}.

\section*{Appendix C: convergence of spectral-Galerkin series}\label{sec:AC}
\begin{figure}
	\centerline{\includegraphics[width=0.8\textwidth]{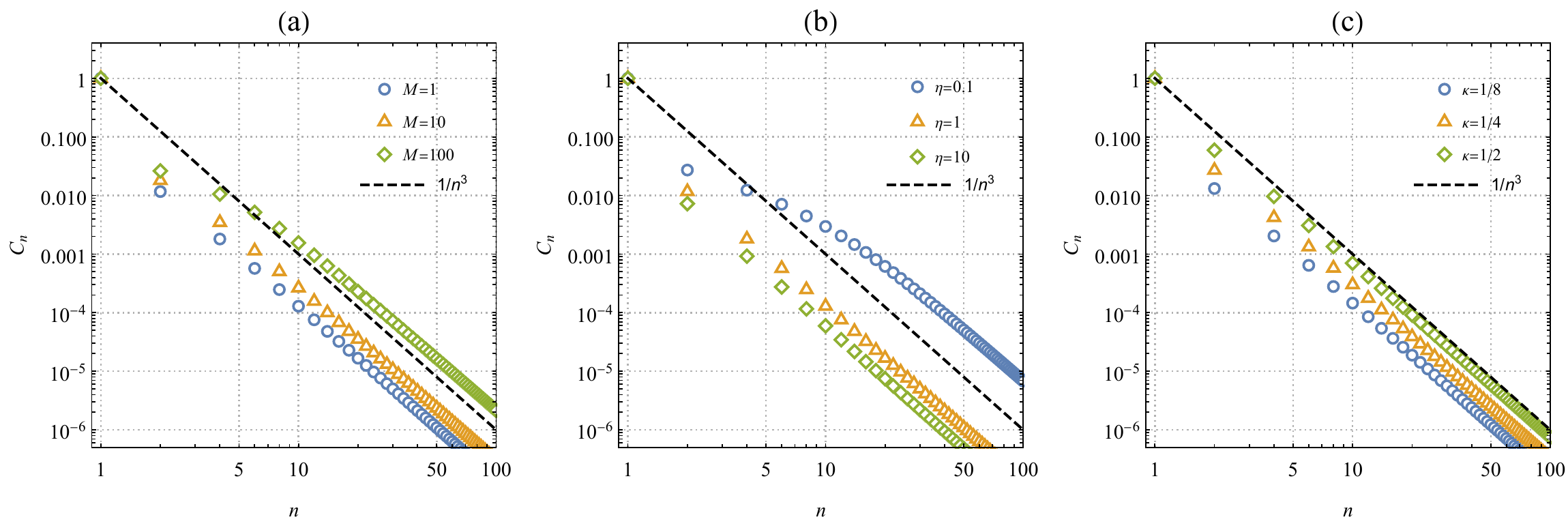}}
	\caption{Magnitude of each term in series \ref{eqn:semi-series} 
		(a) with varying $M$ at $\eta=1$ and $\kappa=1/9$; 
		(b) with varying $\eta$ at $M=1$ and $\kappa=1/9$; 
		(c) with varying $\kappa$ at $M=1$ and $\eta=1$. 
		The dashed line is $1/n^3$}
	\label{fig:con}
\end{figure}

To examine the convergence of this spectral-Galerkin series, we observe the magnitude of each $n^\text{th}$ term in series \ref{eqn:semi-series} and define $C_n$ as
\begin{equation}
	C_n=\frac{||u_{r,n}(r)u_{z,n}(z)||}{||u_{r,1}(r)u_{z,1}(z)||}
\end{equation}
The spectral-Galerkin operation involves $M$, $\eta$ and $\kappa$ in $C_n$.
Figure \ref{fig:con} (a) compares the decline of $C_n$ with various $M$, while (b) and (c) illustrate the effect of geometric parameters $\eta$ and $\kappa$.
It is noticed that although the smaller $M$, $\eta$ and $\kappa$ can lead to smaller $C_n$ ($n>1$), the rate of decay in all the cases levels off to $1/n^3$.
In that context, three important conclusions are reached:
\begin{itemize}
	\item[1.] The nature of convergence of the method is free of Hartmann number $M$, geometric parameters $\eta$ and $\kappa$. In other words, the convergence is \textbf{unconditional}.
	\item[2.] The speed of convergence is $\sim1/n^3$.
	\item[3.] Considering the last item, if series \ref{eqn:semi-series} is truncated by the $N^\text{th}$ term, the round-off error will be at the level of $\sim1/N^2$.
\end{itemize}

\section*{Appendix D: convergence of perturbation expansion}\label{sec:AD}
The applicability of $Re$ rest with the domain of convergence of perturbation expansion\ref{eqn:pertubation}.
Mathematically accuracy convergence domain is beyond the scope of our investigation.
Nevertheless, an estimation of convergence can be fruitful.
In this estimation, we take $\eta\sim1$, $M\gg1$ and $\kappa<1$, which are the common case in the previous study \citep{baylis_mhd_1971,moresco_experimental_2004,zhao_instabilities_2012,poye_scaling_2020}
Scale analysis of equation\ref{eqn:pm2} shows:
\begin{equation}
	w_1\sim\kappa^2u_0^2/M^2
\end{equation}
Similarly equation\ref{eqn:pm2} implicates (note that the boundary layer make $M\partial_{z}h\sim 1$):
\begin{equation}
	u_2\sim\kappa u_0w_1\sim\kappa^3 u_0^3/M^2
\end{equation}
Then, for higher order, $w_2=0$, and
\begin{equation}
	w_3\sim\kappa^2u_0u_2/M^2\sim\kappa^5 u_0^4/M^4
\end{equation}
At this point, it is clear that:
\begin{equation}
	w=Re\sum_{i=1}w_{2i-1}Re^{2i}, \quad w_{2i-1}\sim\kappa^{3i-1} u_0^{2i}/M^{2i}
\end{equation}
Mathematically, the convergence domain shall be approximately:
\begin{equation}
	\frac{Re^{2}u_0^{2}}{M^{2}}<\frac{1}{\kappa^3}
\end{equation}
Moreover, equation\ref{eqn:v:CRZ} implicates that when $M\ll1$
\begin{equation}
	u_0\sim\bar{v}\sim\frac{1}{2} \ln{\left(\frac{1+\kappa}{1-\kappa }\right)}\sim\kappa+O(\kappa^3).
\end{equation}
Thus, the convergence domain is
\begin{equation}
	Re'=\frac{Re}{M}<\frac{1}{\kappa^{2.5}}
\end{equation}
\section*{Acknowledgement}
This work is supported by the National Natural Science Foundation of China (Grant No. 43211872093). We also wish to thank Dr. 	
Alexandre Poyé from CNRS and Prof. Thomas M. York from the Ohio State University for the fruitful discussion.
\section*{Declaration of Interests}
The authors report no conflict of interest.

\bibliographystyle{apalike}
\bibliography{references}  

%
%
%
%

\end{document}